\documentclass[twocolumn]{aastex61}

\shorttitle{Kepler Habitable Zone Exomoons}
\shortauthors{Michelle Hill et al.}
\usepackage{footnote}
\usepackage{float}
\usepackage{graphicx}
\usepackage[noabbrev]{cleveref}

\begin{document}

\title{Exploring Kepler Giant Planets in the Habitable Zone}

\author{Michelle L. Hill}
\affiliation{University of Southern Queensland, Computational Engineering and Science Research Centre, Toowoomba, Queensland 4350, Australia}
\affiliation{University of New England, Department of Physics, Armidale, NSW 2351, Australia}
\affiliation{San Francisco State University, Department of Physics \& Astronomy, 1600 Holloway Avenue, San Francisco, CA 94132, USA}
\email{michellelouhill@gmail.com}

\author{Stephen R. Kane}
\affiliation{University of California, Riverside, Department of Earth Sciences, CA 92521, USA}
\affiliation{University of Southern Queensland, Computational Engineering and Science Research Centre, Toowoomba, Queensland 4350, Australia}

\author{Eduardo Seperuelo Duarte}
\affiliation{Instituto Federal do Rio de Janeiro, Grupo de F\'isica e Astronomia, R. Cel. Delio Menezes Porto, 1045, CEP 26530-060, Nil\'opolis - RJ, Brazil}

\author{Ravi K. Kopparapu}
\affiliation{NASA Goddard Space Flight Center, 8800 Greenbelt Road, Mail Stop 699.0 Building 34, Greenbelt, MD 20771}
\affiliation{NASA Astrobiology Institute’s Virtual Planetary Laboratory, P.O. Box 351580, Seattle, WA 98195, USA}
\affiliation{Sellers Exoplanet Environments Collaboration, NASA Goddard Space Flight Center.}

\author{Dawn M. Gelino}
\affiliation{NASA Exoplanet Science Institute, Caltech, MS 100-22, 770 South Wilson Avenue, Pasadena, CA 91125, USA}
 
\author{Robert A. Wittenmyer}
\affiliation{University of Southern Queensland, Computational Engineering and Science Research Centre, Toowoomba, Queensland 4350, Australia}


\begin{abstract}

The {\it Kepler} mission found hundreds of planet candidates within the Habitable Zones (HZ) of their host star, including over 70 candidates with radii larger than 3 Earth radii ($R_\oplus$) within the optimistic HZ (OHZ) \citep{Kane16}. These giant planets are potential hosts to large terrestrial satellites (or exomoons) which would also exist in the HZ. We calculate the occurrence rates of giant planets ($R_p =$~3.0--25~$R_\oplus$) in the OHZ and find a frequency of $(6.5 \pm 1.9)\%$ for G stars, $(11.5 \pm 3.1)\%$ for K stars, and $(6 \pm 6)\%$ for M stars. We compare this with previously estimated occurrence rates of terrestrial planets in the HZ of G, K and M stars and find that if each giant planet has one large terrestrial moon then these moons are less likely to exist in the HZ than terrestrial planets. However, if each giant planet holds more than one moon, then the occurrence rates of moons in the HZ would be comparable to that of terrestrial planets, and could potentially exceed them. We estimate the mass of each planet candidate using the mass-radius relationship developed by \citet{Chen16}. We calculate the Hill radius of each planet to determine the area of
influence of the planet in which any attached moon may reside, then calculate the estimated angular separation of the moon and planet for future imaging missions. Finally, we estimate the radial velocity semi-amplitudes of each planet for use in follow up observations.

\end{abstract}



\section{Introduction}
\label{introduction}

The search for exoplanets has progressed greatly in the last 3 decades
and the number of confirmed planets continues to grow steadily. These
planets orbiting stars outside our solar system have already provided
clues to many of the questions regarding the origin and prevalence of
life. They have provided further understanding of the formation and
evolution of the planets within our solar system, and influenced an
escalation in the area of research into what constitutes a habitable
planet that could support life. With the launch of NASA's {\it Kepler} telescope thousands of planets were found, in particular planets as far out from their host star as the
Habitable Zone (HZ) of that star were found, the HZ being defined as the region
around a star where water can exist in a liquid state on the surface
of a planet with sufficient atmospheric pressure \citep{kas93}. The 
HZ can further divided into two regions called the 
conservative HZ (CHZ) and the optimistic HZ (OHZ) \citep{Kane16}.
The CHZ inner edge consists of the runaway greenhouse limit, where a chemical breakdown 
of water molecules by photons from the sun will allow the now free hydrogen 
atoms to escape into space, drying out the planet at 0.99~AU 
in our solar system \citep{Kopp14}. The CHZ outer edge consists of the 
maximum greenhouse effect, at 1.7~AU 
in our solar system, where the temperature on the planet drops to a point 
where CO$_2$ will condense permanently, which will in turn increase the planet's 
albedo, thus cooling the planet's surface to a point where all water is frozen 
\citep{Kalt11}. The OHZ in our solar system lies between
0.75--1.8~AU, where the inner edge is the "recent Venus" limit, based
on the empirical observation that the surface of Venus has been dry
for at least a billion years, and the outer edge is the "early Mars"
limit, based on the observation that Mars appears to have been
habitable $\sim$3.8 Gyrs ago \citep{Kopp13}. The positions of the 
HZ boundaries vary in other planetary systems in accordance 
with multiple factors including the effective temperature, stellar flux and luminosity of a host star. 

A primary goal of the {\it Kepler} mission was to determine the
occurrence rate of terrestrial-size planets within the HZ of their
host stars. \citet{Kane16} cataloged all
Kepler candidates that were found in their HZ, providing a list of HZ
exoplanet candidates using the Kepler data release 24, Q1--Q17 data
vetting process, combined with the revised stellar parameters from
DR25 stellar properties table. Planets were then split into 4 groups
depending on their position around their host star and their
radius. Categories 1 and 2 held planets that were $<2$~$R_{\oplus}$ in the
CHZ and OHZ respectively and Categories 3 and 4
held planets of any radius in the CHZ and OHZ
respectively. In Category 4, where candidates of any size radius are
found to be in the OHZ, 76 planets of size 3~$R_\oplus$ and above were found.

Often overshadowed by the discoveries of numerous transiting Earth-size 
planets in recent years \citep[e.g.][]{trappist, small1}, 
Jupiter-like planets are nonetheless a critical feature of a planetary 
system if we are to understand the occurrence of truly Solar-system like 
architectures.  The frequency of close-in planets, with orbits 
$a\leq$0.5~AU, has been investigated in great detail thanks to the 
thousands of {\it Kepler} planets \citep{How12, fressin13, 
Burke15}.  In the icy realm of Jupiter analogs, giant planets in 
orbits beyond the ice line $\sim$3\,AU, radial velocity (RV) legacy surveys 
remain the critical source of insight.  These surveys, with time 
baselines exceeding 15 years, have the sensitivity to reliably detect or 
exclude Jupiter analogs \citep{wittenmyer06, cumming08, jupiters1, 
rowan16}.  For example, an analysis of the 18-year Anglo-Australian 
Planet search by \citet{jupiters2} yielded a Jupiter-analog occurrence 
rate of $6.2^{+2.8}_{-1.6}$\% for giant planets in orbits from 3 to 7 AU.  
Similar studies from the Keck Planet search \citep{cumming08} and the 
ESO planet search programs \citep{zech13} have arrived at 
statistically identical results: in general, Jupiter-like planets in 
Jupiter-like orbits are present around less than 10\% of solar-type 
stars. While these giant planets are not favored in the search for Earth-like planets, 
the discovery of a number of these large planets in the habitable zone 
of their star \citep{Diaz16} do indicate a potential for large rocky moons also residing in the HZ.

A moon is generally defined as a celestial body that orbits around a planet or asteroid and whose orbital barycenter is located inside the surface of the host planet or asteroid. There are currently 175 known satellites orbiting the 8 planets within the solar system, most of which are in orbit around the two largest planets in our system with Jupiter hosting 69 known moons and Saturn hosting 62 known moons\footnote{\tt http://www.dtm.ciw.edu/users/sheppard/satellites/}. The diverse compositions of the satellites in the solar system give insight into their formation \citep{can02,hel15}. Most moons are thought to be formed from accretion within the discs of gas and dust circulating around planets in the early solar system. Through gravitational collisions between the dust, rocks and gas the debris gradually builds, bonding together to form a satellite \citep{Elser11}. Other satellites may have been captured by the gravitational pull of a planet if the satellite passes within the planets area of gravitational influence, or Hill radius. This capture can occur either prior to formation during the proto-planet phase, as proposed in the nebula drag theory \citep{Holt17, Poll79}, or after formation of the planet, also known as dynamical capture. Moons obtained via dynamical capture could have vastly different compositions to the host planet and can explain irregular satellites such as those with high eccentricities, large inclinations, or even retrograde orbits \citep{Holt17, Nes03}. The Giant-Collision formation theory, widely accepted as the theory of the formation of Earth’s Moon, proposes that during formation the large proto-planet of Earth was struck by another proto-planet approximately the size of Mars that was orbiting in close proximity. The collision caused a large debris disk to orbit the Earth and from this the material the Moon was formed \citep{Hart75, Came76}. The close proximity of each proto-planet explains the similarities in the compositions of the Earth and Moon while the impact of large bodies helps explain the above average size of Earth’s Moon \citep{Elser11}. The large number of moons in the solar system, particularly the large number orbiting the Jovian planets, indicate a high probability of moons orbiting giant exoplanets.  

Exomoons have been explored many times in the past
\citep[e.g.][]{Williams97, Kipping09b, Hell12}. Exomoon
habitability particularly has been explored in great detail by Dr Rene
Heller,  \citep[e.g.][]{Hell12, HellBar13, HellPud15, Zoll17} who 
proposed that an exomoon may even provide a
better environment to sustain life than Earth. Exomoons have the
potential to be what he calls "super-habitable" because they offer a
diversity of energy sources to a potential biosphere, not just a
reliance on the energy delivered by a star, like earth. The biosphere
of a super-habitable exomoon could receive energy from the reflected
light and emitted heat of its nearby giant planet or even from the
giant planet's gravitational field through tidal forces. Thus exomoons
should then expect to have a more stable, longer period in which the
energy received could maintain a livable temperate surface condition
for life to form and thrive in.

Another leader in the search for exomoons has been the "Hunt for
Exomoons with Kepler" (HEK) team; \citep[e.g.][]{Kipping12, Kipping13a, Kipping13b, Kipping14, Kipping15}. Here Kipping and others investigated the potential capability and the results of {\it Kepler}, focusing on the use of transit timing variations (TTV's) and and transit duration variations (TDV's) to detect exomoon
signatures. Though several attempts to search for companions to
exoplanets through high-precision space-based photometry
yielded null results, the latest HEK paper \citep{Teachey17} indicates the potential
signature of a planetary companion, exomoon Candidate Kepler-1625b
I. This exomoon is yet to be confirmed and as such caution must be
exercised as the data is based on only 3 planetary transits. 
Still, this is the closest any exomoon hunter has
come to finding the first exomoon. As we await the results of the
follow up observations on this single candidate, it is clear future
instruments will need greater sensitivity for the detection of
exomoons to prosper. While the HEK papers focused on using the TTV/TDV methodology's to
detect exomoons around all of the {\it Kepler} planets, our paper
complements this study by determining the estimated angular separation
of only those Kepler planet candidates $3R_\oplus$ and above that are
found in the optimistic HZ of their star.  We choose the lower limit
of 3$R_\oplus$ as we are interested only in those planets deemed to be
gas giants that have the potential to host large satellites. While
there is a general consensus that the boundary between terrestrial and
gaseous planets likely lies close to 1.6$R_\oplus$, we use 3$R_\oplus$
as our cutoff to account for uncertainties in the stellar and
planetary parameters and prevent the inclusion of potentially
terrestrial planets in our list, as well as planets too small to host
detectable exomoons. We use these giant planets to determine the
future mission capabilities required for imaging of potential HZ
exomoons. We also include RV semi-amplitude calculations
for follow up observations of the HZ giant planets.

In Section \ref{SM} of this paper we explore the potential of these HZ
moons, citing the vast diversity of moons within our solar
system. We predict the frequency of HZ giant planets using the
inverse-detection-efficiency method in Section~\ref{hzgiant}. In Section~\ref{propHZgiant} we present the calculations and results for the estimated planet mass, Hill radius of the planet, angular separation of the planet from the
host star and of any potential exomoon from its host planet, and the RV semi-amplitude of the planet on its host
star. Finally, in Section~\ref{Xmoon} we discuss the
calculations and their implications for exomoons and outline 
proposals for observational prospects of the planets and
potential moons, providing discussion
of caveats and concluding remarks.


\section{Science Motivation}
\label{SM}

Within our solar system we observe a large variability of moons in
terms of size, mass, and composition. Five icy moons of Jupiter and
Saturn show strong evidence of oceans beneath their surfaces:
Ganymede, Europa and Callisto at Jupiter, and Enceladus and Titan at
Saturn. From the detection of water geysers and deep oceans below the icy crust of Enceladus
\citep{Porco06, Hsu15} to the volcanism on Io \citep{Morabito79}, our own solar
system moons display a diversity of geological phenomena and are
examples of potentially life holding worlds. Indeed Ganymede, the
largest moon in our solar system, has its own magnetic field
\citep{Kivelson96}, an attribute that would increase the potential
habitability of a moon due to the extra protection of the moons
atmosphere from its host planet \citep{Williams97}. And while the
moons within our own HZ have shown no signs of life, namely Earth's
moon and the Martian moons of Phobos and Deimos, there is still great
habitability potential for the moons of giant exoplanets residing in
their HZ.

The occurrence rate of moons in the HZ is intrinsically connected to
the occurrence rate of giant planets in that region. We thus consider
the frequency of giant planets within the OHZ. We choose to use the wider OHZ due to warming effects any exomoon will
undergo as it orbits its host planet. The giant planet will increase the
effective temperature of the moon due to contributions of
thermal and reflected radiation from the giant planet
\citep{Hinkel13}. Tidal effects will also play a significant role, as
seen with Io. \citet{Scharf06} proposed that this heating mechanism
can effectively increase the outer range of the HZ for a moon as the
extra mechanical heating can compensate for the lack of radiative
heating provided to the moon. For the same reason this could reduce
the interior edge of the HZ causing any moon with surface water to
undergo the runaway green house effect earlier than a lone body
otherwise would, though the outwards movement of the inner edge has 
been found to be significantly less than that of the outer edge and so the 
effective habitable zone would still be widened for any exomoon. 
This variation could also possibly enable giant
exoplanets with eccentric orbits that lie, at times, outside the OHZ
to maintain habitable conditions on any connected exomoons
\citep{Hinkel13}.


\section{Frequency of Habitable Zone Giant Planets}
\label{hzgiant}

The occurrence rates of terrestrial planets in the HZ has been explored many times in the literature \citep[e.g.][]{How12, Dress13, Dress15, Kopp13a, Peti13}. The planet occurrence rate is defined as the number of planets per star
(NPPS) given a range of planetary radius and orbital period. It is
simply represented by the expression
\begin{equation}
NPPS = \frac{N_p}{N_*}
\end{equation}
where $N_p$ is the real number of planets and $N_*$ is the number of
stars in the {\it Kepler} survey. However, $N_p$ is unknown due to some
limitations of the mission. The first limitation is produced by the
duty cycle which is the fraction of time in which a target was
effectively observed \citep{Burke15}. The requirement adopted by the
Kepler mission to reliably detect a planet is to observe at least
three consecutive transits \citep{Koch10}. This requirement is
difficult to achieve for low duty cycles and for planets with long
orbital periods. The second limitation is the photometric efficiency,
the capability of the photometer to detect a transit signal for a
given noise (Signal-to-Noise ratio; SNR). For a given star it is
strongly dependent on the planet size since the transit depth depends
on the square of the radius ratio between the planet and the
star. Thus, smaller planets are more difficult to detect than the
bigger ones. Finally, the transit method is limited to orbits nearly
edge-on relative to the telescope line of sight. Assuming a randomly
oriented circular orbit, the probability of observing a star with
radius $R_*$ being transited by a planet with semi-major axis $a$ is
given by $R_*/a$.

Those survey features contribute to the underestimation of the number of
detectable planets orbiting the stars of the survey. Thus, to obtain
$N_p$, the observed number of planets $N_{obs}$ is corrected by taking
the detection efficiencies described above into account. In Section \ref{method}, the method used to accomplish this goal is described.

\subsection{The Method}
\label{method}

The method used in this work to compute the occurrence rate, which is
commonly used in the literature (\citep{How12}, \citep{Dress15}), is
called the inverse-detection-efficiency method \citep{Fore16}. It
consists of calculating the occurrence rates in a diagram of radius
and period binned by a grid of cells. The diagram is binned following
the recommendations of the NASA ExoPAG Study Analysis Group 13, i.e,
the i-th,j-th bin is defined as the interval $ [1.5^{i-2}, 1.5^{i-1})R_\oplus$ and 10x$[2^{j-1}, 2^{j})day $. The candidates are plotted, according to their physical parameters, and the real number of planets is then computed in each
cell ($N_p^{i,j}$) by summing the observed planets
($N_{obs}^{i,j}$) in the i,j bin weighted by their inverse
detection probability, as
\begin{equation}
	N_p^{i,j} = \sum _{n=1}^{N_{obs}^{i,j}} \frac{1}{p_n}
\end{equation}
where $p_n$ is the detection probability of planet $n$. Finally, the
occurrence rate is calculated by Equation (3) as a function of orbital
period and planetary radius,
\begin{equation}
	NPPS^{i,j} = \frac{N_p^{i,j}}{N_*}
\end{equation}

\subsection{Validating Methodology}
\label{valid} 

We confirm that we are able to recover accurate occurrence rates by
using the method described above to first compute the occurrence rates
of planets orbiting M dwarfs and comparing the results with
known values found by \citep{Dress15} (here after DC15). DC15
used a stellar sample of 2543 stars with effective
temperatures in the range of 2661--3999~K, stellar radii between
0.10 and 0.64~$R_\oplus$, metallicity spanning from -2.5 to 0.56 and
Kepler magnitudes between 10.07 and 16.3 \citep{Burke15}. The sample contained 156 candidates with orbital periods extending from 0.45 to 236 days and planet radii from 0.46 to 11$ R_\oplus$.

The real number of planets was computed in each cell using equation (2) with $p_n$ being the average detection probability
of planet $n$. Then equation (3) was used to calculate the occurrence
rates considering the real number of planets and the total number of
stars used in the sample. We then recalculated the occurrences using the candidates from DC15 but with their disposition scores and planetary radius updated by the NASA Exoplanet Archive \citep{Akeson13}. The disposition score is a value between 0 and 1 that indicates the confidence in the KOI disposition, a higher value indicates more confidence in its disposition. The value is calculated from a Monte Carlo technique such that the score's value is equivalent to the fraction of iterations where the Robovetter yields a disposition of "Candidate" \citep{Akeson13}. From the 156 candidates used by DC15, 28 candidates were removed from the sample because their disposition had changed in the NASA Exoplanet Archive. 

We found there is a good agreement between the results obtained in this work and those obtained
by DC15 in the smaller planets domain, 
particularly in the range of 1.5--3.0~$R_\oplus$, while the occurrence rates for larger planets tended to be smaller in this work than the DC15 results. 
As our method validation compared the occurrence rates results obtained
by two works that utilize basically the same method, data and
planetary physical parameters, the discrepancies we observed may have been
produced by differences in the detection probabilities used.

\subsection{Stellar Sample}
\label{sample}

\begin{figure*}[t!]
  \includegraphics[width=\linewidth]{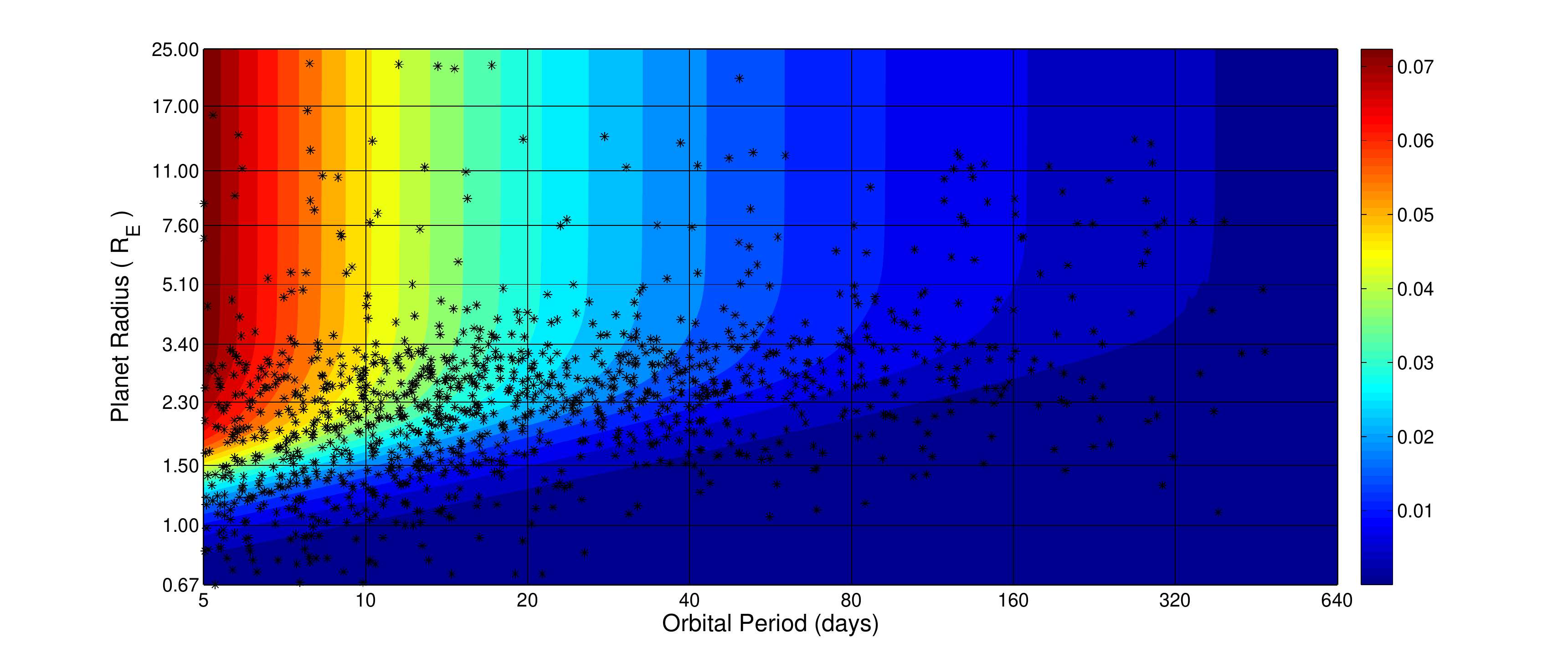}
  \caption{Average detection probability for G stars as a function of
    planet radius and orbital period. The star symbols represent the 1,819
    {\it Kepler} candidates detected for these stars. Note the color bar to the right indicates the detection probability of the planets with greatest probability of detection corresponding with the top of the scale. Planets found on the top left corner of the graph will have a greater probability of detection.}
  \label{fig:NPPScand}
\end{figure*}

We selected a sample of 99,417 stars with $2400~K \leq T_\mathrm{eff} < 6000$~K and $\log g \geq 4.0$ from the Q1--17 Kepler Stellar Catalog in the
NASA Exoplanet Archive. From those stars, 86,383 stars have detection
probabilities computed in the range of 0.6--25~$R_\oplus$ and 5--700~days (Burke, private communication). The average detection
probability was calculated for each G, K and M stars subsample and
then used to compute the occurrence rates as a function of spectral
type as described in Section 3.1. The number of stars in each spectral type category are shown in Table~\ref{occrategiants}, where the properties of the stars in each category follow the prescription of the NASA ExoPAG Study Analysis Group 13. Figure
\ref{fig:NPPScand} shows the diagram divided into cells which are
superimposed by the average detection probability for G stars.

\subsection{Planet Candidates Properties}
\label{properties}

The properties of all 4034 candidates/confirmed planets were
downloaded from the Q1--17 Kepler Object of Interest on the NASA
Exoplanet Archive. From this we selected 2,586 candidates that orbit
the sample of stars described in the previous section and whose
planetary properties lie inside the range of parameters in which the
detection efficiencies were calculated. We took a conservative approach and discarded candidates
with disposition scores smaller than 0.9. The
properties of the resulting candidate sample range from 0.67--22.7~$R_\oplus$
and from 5.0--470~day orbits. The planetary sample was divided into subsamples according to the spectral type of their host
stars, leaving us with 1207 planets orbiting G stars, 534 planets
orbiting K stars and 93 planets orbiting M stars.

\subsection{Planet Occurrence Rates}
\label{occRates}

\begin{figure}
\begin{center}
\includegraphics[width=\columnwidth]{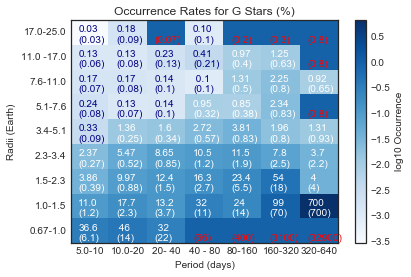}
\caption{Binned planet occurrence rates for G stars as a function of
  planet radius and orbital period. Planet occurrence is given as a
  percentage along with uncertainty percentage (in brackets). For bins without planets we compute the uncertainty, and thus upper limit by including one detection at the center of the bin. The bins treated this way have been colored with red font for transparency.}
\label{fig:Goccur}
\end{center}
\end{figure}

\begin{figure}
\begin{center}
\includegraphics[width=\columnwidth]{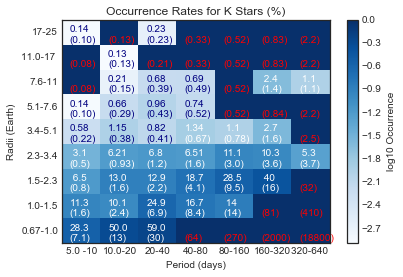}
\caption{Binned planet occurrence rates for K stars as a function of
  planet radius and orbital period. Planet occurrence is given as a
  percentage along with uncertainty percentage (in brackets). For bins without planets we compute the uncertainty, and thus upper limit by including one detection at the center of the bin. The bins treated this way have been colored with red font for transparency.}
\label{fig:Koccur}
\end{center}
\end{figure}

\begin{figure}
\begin{center}
\includegraphics[width=\columnwidth]{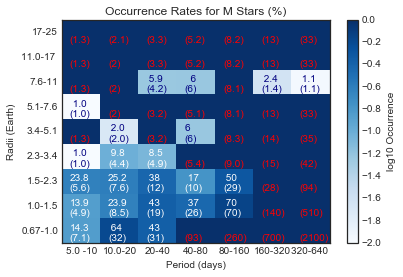}
\caption{Binned planet occurrence rates for M stars as a function of
  planet radius and orbital period. Planet occurrence is given as a
  percentage along with uncertainty percentage (in brackets). For bins without planets we compute the uncertainty, and thus upper limit by including one detection at the center of the bin. The bins treated this way have been colored with red font for transparency.}
\label{fig:Moccur}
\end{center}
\end{figure}

For each sample of spectral type, the occurrence rates were computed
for each cell spanning a range of planet radius and orbital period
following the method described in Section 3.1 and using equation
2. For those cells in which no candidate was observed, we estimated an upper limit based on the uncertainty of the occurrence rate as if there was one detection in the center of the bin. Figures
\ref{fig:Goccur}, \ref{fig:Koccur} and \ref{fig:Moccur} show the occurrence rates for
each cell. The uncertainties were estimated using the relation
\begin{equation}
\delta NPPS^{i,j} = \frac{NPPS^{i,j}}{\surd N_p^{i,j}}
\end{equation}

\subsection{Frequency versus Planet Radius and Insolation}
\label{freq}

\begin{figure*}[t!]
\includegraphics[width=\linewidth]{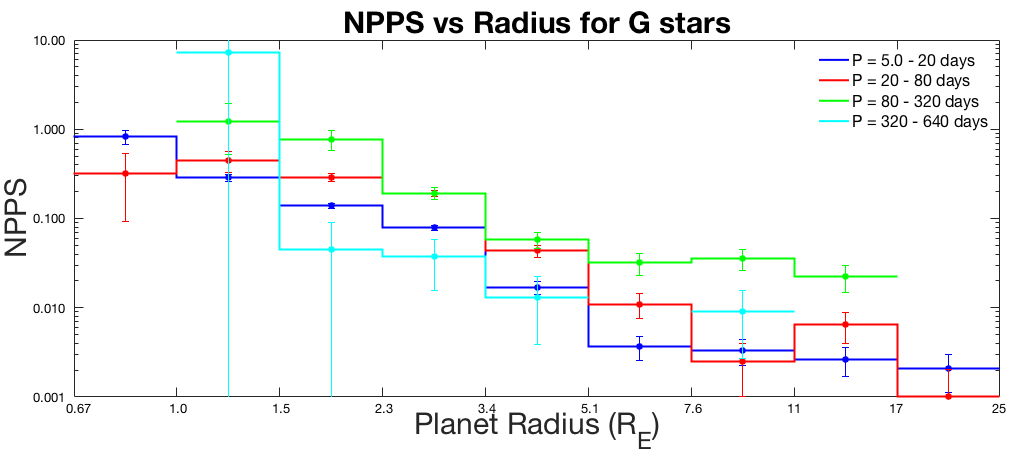}
\caption{Number of Planets Per Star (NPPS) vs radius for G stars. Each line color represents a set range of periods. The data indicates that, for G stars, planets with radii greater than 1.5~$R_\oplus$ are most commonly found with orbital periods between 80--320~days. Also the occurrence rate of planets with orbits between 320--640 days shows a large spike for planets with radii between 1.0--1.5~$R_\oplus$.}
\label{fig:NPPSRAD1}
\end{figure*}

\begin{figure*}[t!]
\includegraphics[width=\linewidth]{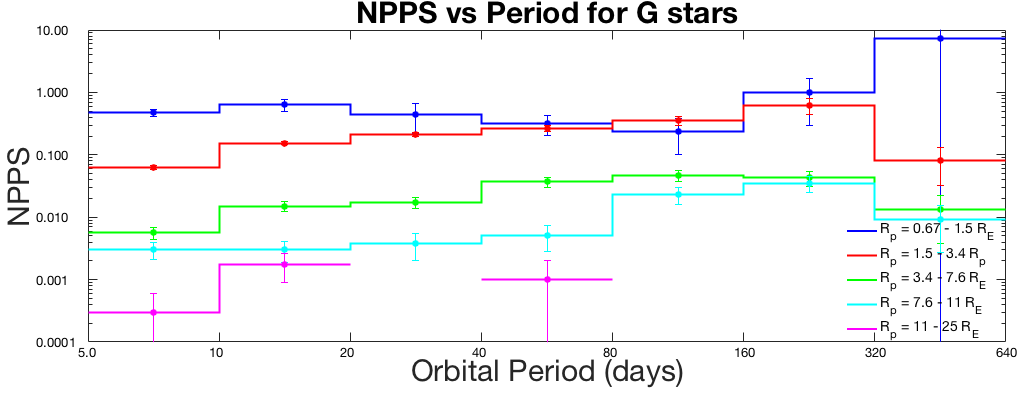}
\caption{Number of Planets Per Star (NPPS) vs period for G stars. Each line color represents a set range of radii. The data indicates that, for G stars, small planets are more abundant than giant planets in each orbital period bin. The magenta line indicating planets with radii between 11 and 25~$R_\oplus$ represents the rarest objects detected by {\it Kepler}, thus there is a lack of sufficient data to complete the calculations of their occurrence rates at longer orbital periods.}
\label{fig:NPPSper1}
\end{figure*}

\begin{figure*}[t!]
\includegraphics[width=\linewidth]{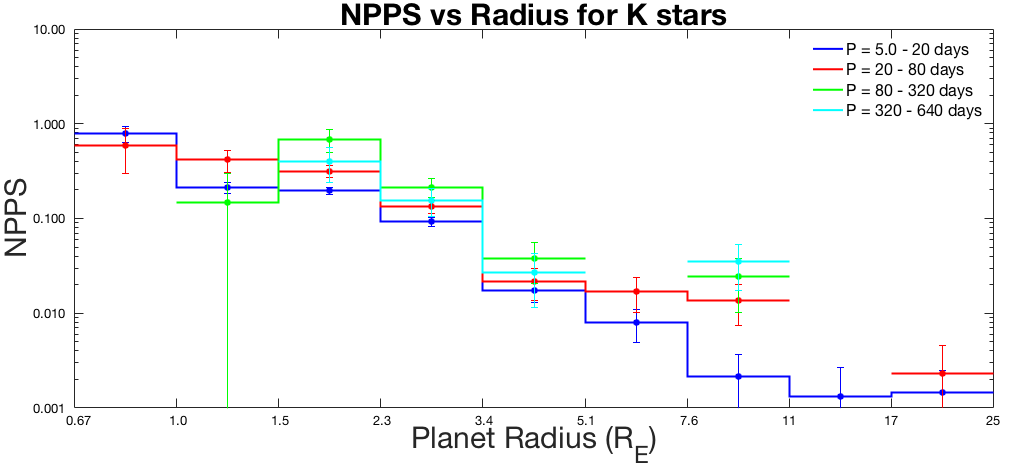}
\caption{Number of Planets Per Star (NPPS) vs radius for K stars. Each line color represents a set range of periods. The data indicates that planets with radii between 1.5--5.1~$R_\oplus$ most commonly have orbital periods between 80--320 days. Also, for K stars, small planets are more abundant than giant planets in each orbital period bin.}
\label{fig:NPPSRad2}
\end{figure*}

\begin{figure*}[t!]
\includegraphics[width=\linewidth]{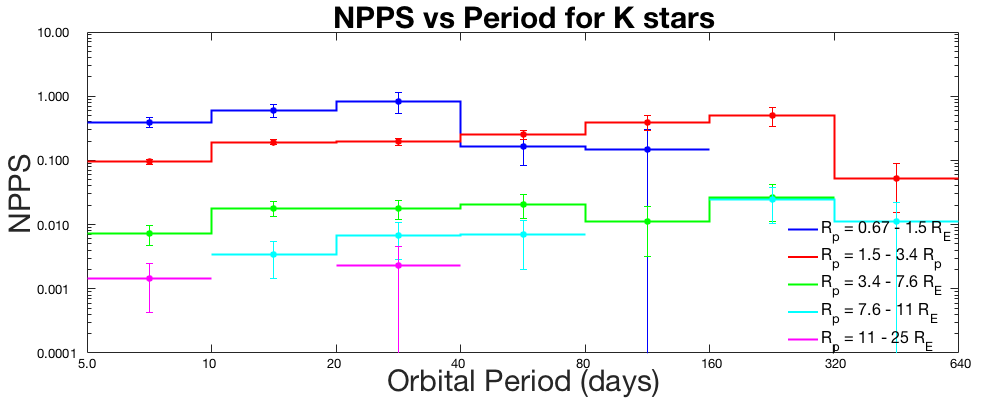}
\caption{Number of Planets Per Star (NPPS) vs period for K stars. Each line color represents a set range of radii. Note there is a drop in the blue line representing the lowest mass planets between 0.67--1.5~$R_\oplus$ at an orbital period of 40 days. This corresponds to the limit of detection efficiency of {\it Kepler} for small planets and thus there is not sufficient data in this region to claim that this is a significant drop.}
\label{fig:NPPSper2}
\end{figure*}

\begin{figure*}[t!]
\includegraphics[width=\linewidth]{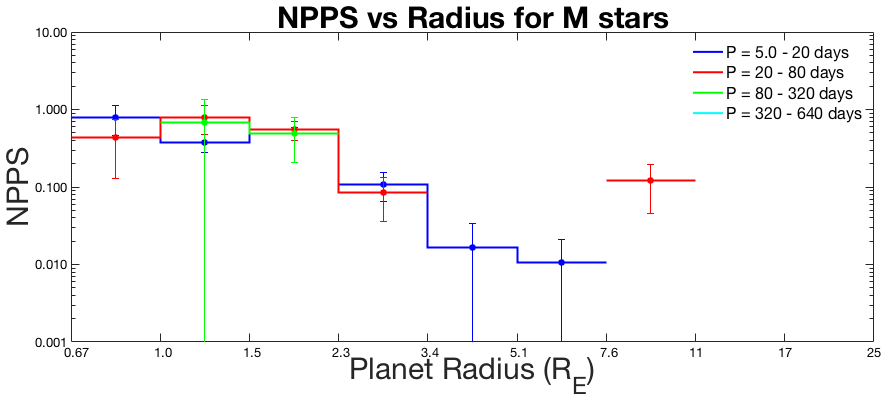}
\caption{Number of Planets Per Star (NPPS) vs radius for M stars. Each line color represents a set range of periods. We observe a lack of any planets with $R_p > 11$~$R_\oplus$. Planets with $R_p =$~7.6--11~$R_\oplus$ tend to be found with orbital periods between 20--80 days.}
\label{fig:NPPSRad3}
\end{figure*}

\begin{figure*}[t!]
\includegraphics[width=\linewidth]{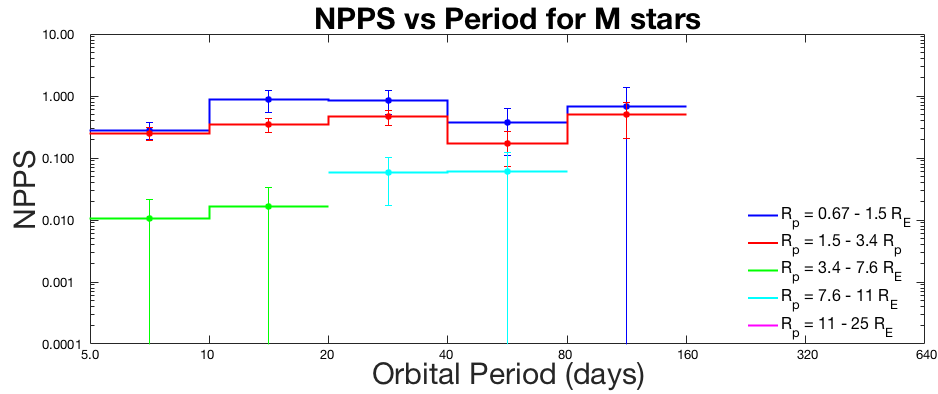}
\caption{Number of Planets Per Star (NPPS) vs period for M stars. Each line color represents a set range of radii. We observe that small planets tend to be more abundant than giant planets in each orbital period bin. Note the drop in planets beyond an orbital period of 160 days corresponds with the limit of {\it Kepler} detection efficiency for these dim stars.}
\label{fig:NPPSper3}
\end{figure*}

Figure \ref{fig:NPPSRAD1} $-$ \ref{fig:NPPSper3} show the occurrence
rates as a function of planet radius and orbital period.  Figure
\ref{fig:NPPSRAD1} shows the occurrence rates for planets around G
stars.  Number of Planets Per Star (NPPS) is plotted against the
planet radius and each line represents a band of orbital periods. The
data indicates that, for G stars, planets with radii greater than 1.5~$R_\oplus$
are most commonly found with orbital periods between 80-320 days. The
occurrence for planets with orbits between 320-640 days shows a spike
for planets with radii between 1.0--1.5~$R_\oplus$. In general, our results
show that small planets are more abundant than giant planets in each
orbital period bin which is consistent with \citet{witt11, Kane16}.

The trends observed for K stars follows that observed for G stars;
small planets are more abundant than giant planets in each orbital
period bin. While Figure \ref{fig:NPPSper2} shows a complete lack
of giant planets $>11$~$R_\oplus$ with orbital periods $>40$~days, this radius range represents the rarest objects detected by {\it Kepler}, thus there is a lack of sufficient data to complete the calculations of their occurrence rates. In
addition, there appears to be a lack of planets with radius 5.1--7.6~$R_\oplus$
with orbits of $>80$~days.

For M stars, the occurrences for different orbital periods are very
similar. We observe a lack of any giant planets with $R_p > 11$~$R_\oplus$
(Figure~\ref{fig:NPPSRad3}). Planets with $R_p =$~7.6--11~$R_\oplus$
tend to be found with orbital periods between 20--80~days.

\subsection{Frequency of Giants in the Habitable Zone}
\label{freqHZ}

The OHZ for each host candidate was computed following the model described
by \citet{Kopp13,Kopp14}. From the sample of candidates selected and
described in Section 3.3, 12 candidates orbit within the OHZ of their
respective G host stars, 14 candidates orbit in the OHZ of their K host
stars and only 1 candidate orbits in the OHZ of an M star. The properties of the spectral type bins and the occurrence rates of giant planets in the OHZ is shown in Table~\ref{occrategiants}.

\begin{deluxetable}{lcccc}
\centering
\caption{Planet Occurrence rates of giant planets $>3$~$R_\oplus$ in the OHZ of their star.
\label{occrategiants}}
\tablehead{
\colhead{Spectral Type} & \colhead{$T_\mathrm{eff}$ (K)} & \colhead{No. stars} & \colhead{Planets in OHZ} & \colhead{NPPS (\%)}
}
\startdata
G & 5300--6000 & 59510 & 12 &  $6.5 \pm 1.9$ \\
K & 3900--5300 & 24560 & 14 & $11.5 \pm 3.1$ \\
M & 2400--3900 &  2313 &  1 &  $6.0 \pm 6.0$ \\
\enddata
\end{deluxetable}


\section{Properties of Habitable Zone Giant Planets}
\label{propHZgiant}

Here we present the calculations for the estimated planet mass, Hill
radius of the planet, angular separation of the planet from the host
star and of any potential exomoon from its host planet, both estimates
of which can be used in deciding the ideal candidates for future
imaging missions, and finally the RV semi-amplitude of
the planet on its host star for use in follow up observations of each
giant planet.

We start by estimating the mass of each of the Kepler candidates using
the mass/radius relation found in \citet{Chen16}:
\begin{equation}
R_p = M_p^{0.59}
\end{equation}
where $R_p$ is the planet radius in Earth radii and $M_p$ is planet mass in Earth masses. 

As is noted in \citet{Chen16}, this relationship is only reliable up to $\sim 10R_\oplus$. As planets $10R_\oplus$ and above can vary greatly in density and thus greatly in mass, we have chosen to quantify each exoplanet with a radius of $10R_\oplus$ or greater as 3 set masses; 1 Saturn mass for the very low density planets, 1 Jupiter mass for a direct comparison with our solar system body, and 13 Jupiter mass for the higher density planets. As there is discrepancy as to the mass of a planet vs brown dwarf we have chosen to use the upper limit of 13 Jupiter masses. For any planet found to have a mass larger than this the Hill radius and RV signal will thus be greater than that calculated. 

Using our mass estimate, we first consider the radius at which a moon is gravitationally bound to a planet, calculating the Hill radius using \citet{Hinkel13}:
\begin{equation}
r_H = a_{sp} \chi (1-e_{sp}) \left( \frac{M_p}{M_\star} \right)^\frac{1}{3}
\end{equation}
where $M_\star$ is the mass of the host star. Assuming an eccentricity of the planet--star system of $e = 0$, the above equation becomes: 
\begin{equation}
	r_H = a_{sp} \chi \left( \frac{M_p}{M_\star} \right)^\frac{1}{3}
\end{equation}
The factor $\chi$ is added to take into account the fact that the Hill radius is just an estimate. Other effects may impact the gravitational stability of the system, so following \citep{Barnes02}, \citep{Kipping09a} and \citep{Hinkel13}, we have chosen to use a conservative estimate of $ \chi \leq 1/3$.

The expected angular separation of the exomoon for its host planet is then calculated by: 
\begin{equation}
\alpha \arcsec  = \frac{r_H(\chi = {1}/{3})}{d} 
\end{equation} 
Here $d$ represents the distance of the star planet system in parsecs (PC) and Hill radius is expressed in (AU).

Finally, we calculate the RV semi-amplitude, $K$, of each planet given its estimated mass:
\begin{equation}
K = \frac{(2 \pi G)}{P^{{1}/{3}}} \frac{(M_p \sin{i})}{((M_\star + M_p)^{{2}/{3}}}
\end{equation} 
We further assume an orbital inclination of $\sim$90$\degr$ and $e = 0$.

Table \ref{c4HZ} includes each of the parameters used in our calculations which have been extracted from the HZ catalogue \citep{Kane16} as well as the NASA exoplanet archive. Table \ref{c4HRAS} presents our calculations of planet mass, Hill radii, estimated RV semi-amplitudes and angular separations of the planet -- star systems and potential planet -- moon systems at both the full Hill radii (HR) and $\onethird$ Hill radii (\onethird HR).

Tables \ref{c4RVSA} and \ref{c4HR} then present our calculations of Hill radii, angular separations of a potential planet--moon systems at the full Hill radius and RV semi-amplitudes for each exoplanet with a radius of $10R_\oplus$ or greater with our chosen quantified masses; 1 Saturn mass ($M_{sat}$), 1 Jupiter mass ($M_J$), and 13 Jupiter masses (13$M_J$).
    
\floattable
\startlongtable
\begin{deluxetable*}{llllllllll}
  \tablewidth{0pc}
 \tabletypesize{\scriptsize}
  \centering
  \tablecaption{Habitable Zone candidates with $R_p > 3 \ R_\oplus$.
  \label{c4HZ}}
\tablehead{
\colhead{KOI name}  & \colhead{Kepler} & \colhead{$T_\mathrm{eff}$} &  \colhead{Period}      &  \colhead{$a$ \tablenotemark{*}}     &  \colhead{Planet Radius} &  \colhead{Incident Flux} &  \colhead{Stellar Mass} &  \colhead{Distance}    &  \colhead{Magnitude} \\
\colhead{} & \colhead{} & \colhead{K} & \colhead{days}  & \colhead{AU} & \colhead{$R_\oplus$} & \colhead{$F_\oplus$} & \colhead{$M_\star$} & \colhead{PC}  & \colhead{Kepler Band}}
\startdata
K03086.01 & $-$ & $5201\pm83$ & $174.732\pm0.003$ & 0.573 & $3\pm0.235$ & $1.61\pm0.35$ & $0.82\pm0.05$ & $1006\pm84$ & 15.71 \\
K06786.01 & $-$ & $5883\pm186$ & $455.624\pm0.026$ & 1.153 & $3\pm0.585$ & $0.64\pm0.33$ & $0.99\pm0.13$ & $3192\pm550$ & 11.97 \\
K02691.01 & $-$ & $4735\pm170$ & $97.446\pm0$ & 0.373 & $3.05\pm0.265$ & $1.53\pm0.49$ & $0.73\pm0.07$ & $447\pm50$ & 14.98 \\
K01581.02 & 896b & $5510\pm158$ & $144.552\pm0.003$ & 0.516 & $3.06\pm0.475$ & $2\pm0.85$ & $0.88\pm0.09$ & $926\pm170$ & 15.48 \\
K08156.01 & $-$ & $6429\pm182$ & $364.982\pm0.011$ & 1.048 & $3.12\pm0.69$ & $1.74\pm0.96$ & $1.15\pm0.16$ & $978\pm240$ & 14.32 \\
K07700.01 & $-$ & $6382\pm180$ & $631.569\pm0.013$ & 1.491 & $3.13\pm0.655$ & $0.75\pm0.4$ & $1.1\pm0.15$ & $798\pm177$ & 14.00 \\
K04016.01 & 1540b & $4641\pm79$ & $125.413\pm0$ & 0.443 & $3.14\pm0.125$ & $1.19\pm0.18$ & $0.73\pm0.04$ & $293\pm18$ & 14.07 \\
K05706.01 & 1636b & $5977\pm201$ & $425.484\pm0.009$ & 1.155 & $3.2\pm0.61$ & $0.9\pm0.46$ & $1.13\pm0.13$ & $1589\pm348$ & 15.81 \\
K02210.02 & 1143c & $4895\pm78$ & $210.631\pm0.002$ & 0.648 & $3.23\pm0.15$ & $0.71\pm0.11$ & $0.82\pm0.04$ & $607\pm38$ & 15.20 \\
K08276.01 & $-$ & $6551\pm183$ & $385.859\pm0.005$ & 1.107 & $3.23\pm0.705$ & $1.93\pm1.05$ & $1.22\pm0.17$ & $944\pm216$ & 13.99 \\
K04121.01 & 1554b & $5275\pm83$ & $198.089\pm0.002$ & 0.631 & $3.24\pm0.36$ & $1.64\pm0.47$ & $0.86\pm0.05$ & $1164\pm143$ & 15.72 \\
K05622.01 & 1635b & $5474\pm158$ & $469.613\pm0.014$ & 1.117 & $3.24\pm0.46$ & $0.38\pm0.15$ & $0.85\pm0.09$ & $944\pm160$ & 15.70 \\
K07982.01 & $-$ & $6231\pm207$ & $376.38\pm0.047$ & 1.029 & $3.26\pm0.665$ & $1.17\pm0.63$ & $1.03\pm0.13$ & $1436\pm333$ & 15.63 \\
K03946.01 & 1533b & $6325\pm79$ & $308.544\pm0.002$ & 0.963 & $3.28\pm0.565$ & $2.82\pm1.12$ & $1.25\pm0.11$ & $734\pm119$ & 13.22 \\
K08232.01 & $-$ & $5573\pm174$ & $189.184\pm0.004$ & 0.610 & $3.31\pm0.77$ & $2.24\pm1.32$ & $0.85\pm0.1$ & $865\pm212$ & 15.05 \\
K05625.01 & $-$ & $5197\pm181$ & $116.454\pm0.002$ & 0.414 & $3.33\pm0.375$ & $2.07\pm0.75$ & $0.7\pm0.07$ & $894\pm132$ & 16.02 \\
K02073.01 & 357d & $5036\pm200$ & $49.5\pm0$ & 0.246 & $3.43\pm2.04$ & $6.57\pm8.8$ & $0.79\pm0.04$ & $771\pm51$ & 15.57 \\
K02686.01 & $-$ & $4658\pm93$ & $211.033\pm0.001$ & 0.627 & $3.43\pm0.17$ & $0.51\pm0.09$ & $0.74\pm0.04$ & $267\pm17$ & 13.86 \\
K01855.01 & $-$ & $4338\pm125$ & $58.43\pm0$ & 0.248 & $3.45\pm0.3$ & $1.92\pm0.55$ & $0.59\pm0.06$ & $298\pm33$ & 14.78 \\
K02828.02 & $-$ & $4817\pm176$ & $505.463\pm0.008$ & 1.153 & $3.46\pm0.315$ & $0.25\pm0.08$ & $0.8\pm0.05$ & $769\pm95$ & 15.77 \\
K02926.05 & $-$ & $3891\pm78$ & $75.731\pm0.002$ & 0.297 & $3.47\pm0.19$ & $0.74\pm0.14$ & $0.61\pm0.03$ & $425\pm35$ & 16.28 \\
K08286.01 & $-$ & $5440\pm180$ & $191.037\pm0.013$ & 0.634 & $3.54\pm0.6$ & $1.59\pm0.75$ & $0.93\pm0.09$ & $1654\pm335$ & 16.65 \\
K01830.02 & 967c & $5180\pm103$ & $198.711\pm0.001$ & 0.625 & $3.56\pm0.215$ & $1.06\pm0.21$ & $0.83\pm0.05$ & $502\pm37$ & 14.44 \\
K00951.02 & 258c & $4942\pm200$ & $33.653\pm0$ & 0.193 & $3.61\pm2.43$ & $12.16\pm18.1$ & $0.83\pm0.05$ & $1542\pm431$ & 15.22 \\
K01986.01 & 1038b & $5159\pm82$ & $148.46\pm0.001$ & 0.524 & $3.61\pm0.205$ & $1.56\pm0.28$ & $0.87\pm0.04$ & $606\pm42$ & 14.84 \\
K01527.01 & $-$ & $5401\pm107$ & $192.667\pm0.001$ & 0.622 & $3.64\pm0.32$ & $1.52\pm0.39$ & $0.86\pm0.05$ & $743\pm71$ & 14.88 \\
K05790.01 & $-$ & $4899\pm82$ & $178.267\pm0.003$ & 0.571 & $3.71\pm0.21$ & $0.81\pm0.14$ & $0.82\pm0.04$ & $643\pm44$ & 15.52 \\
K08193.01 & $-$ & $5570\pm158$ & $367.948\pm0.005$ & 0.996 & $3.72\pm0.6$ & $0.64\pm0.28$ & $0.97\pm0.09$ & $1116\pm202$ & 15.72 \\
K08275.01 & $-$ & $5289\pm176$ & $389.876\pm0.007$ & 1.002 & $3.76\pm0.46$ & $0.44\pm0.17$ & $0.89\pm0.08$ & $975\pm152$ & 15.95 \\
K01070.02 & 266c & $5885\pm250$ & $107.724\pm0.002$ & 0.457 & $3.89\pm1.89$ & $5.47\pm6.24$ & $0.95\pm0.06$ & $1562\pm280$ & 15.59 \\
K07847.01 & $-$ & $6098\pm217$ & $399.376\pm0.069$ & 1.103 & $3.93\pm1.225$ & $2.67\pm2.04$ & $1.12\pm0.17$ & $2190\pm713$ & 13.28 \\
K00401.02 & 149d & $5381\pm100$ & $160.018\pm0.001$ & 0.571 & $3.96\pm0.68$ & $2.08\pm0.77$ & $0.93\pm0.05$ & $541\pm56$ & 14.00 \\
K01707.02 & 315c & $5796\pm108$ & $265.469\pm0.006$ & 0.791 & $4.15\pm0.96$ & $1.75\pm0.8$ & $0.88\pm0.06$ & $1083\pm147$ & 15.32 \\
K05581.01 & 1634b & $5636\pm171$ & $374.878\pm0.008$ & 1.053 & $4.27\pm1.125$ & $1.5\pm0.97$ & $1.1\pm0.13$ & $1019\pm272$ & 14.51 \\
K01258.03 & $-$ & $5717\pm165$ & $148.272\pm0.001$ & 0.546 & $4.3\pm0.75$ & $2.52\pm1.16$ & $0.98\pm0.11$ & $1217\pm245$ & 15.77 \\
K02683.01 & $-$ & $5613\pm152$ & $126.445\pm0$ & 0.473 & $4.49\pm0.635$ & $2.52\pm0.99$ & $0.89\pm0.1$ & $947\pm147$ & 15.50 \\
K00881.02 & 712c & $5067\pm102$ & $226.89\pm0.001$ & 0.673 & $4.53\pm0.26$ & $0.73\pm0.14$ & $0.79\pm0.04$ & $854\pm59$ & 15.86 \\
K01429.01 & $-$ & $5644\pm80$ & $205.913\pm0.001$ & 0.679 & $4.68\pm0.5$ & $1.86\pm0.5$ & $0.98\pm0.06$ & $1232\pm135$ & 15.53 \\
K00902.01 & $-$ & $3960\pm124$ & $83.925\pm0$ & 0.303 & $4.78\pm0.405$ & $0.62\pm0.18$ & $0.53\pm0.04$ & $348\pm43$ & 15.75 \\
K05929.01 & $-$ & $5830\pm158$ & $466.003\pm0.003$ & 1.165 & $4.92\pm0.875$ & $0.59\pm0.27$ & $0.97\pm0.12$ & $780\pm168$ & 14.69 \\
K00179.02 & 458b & $6226\pm118$ & $572.377\pm0.006$ & 1.406 & $5.8\pm0.905$ & $1.15\pm0.45$ & $1.13\pm0.09$ & $904\pm140$ & 13.96 \\
K03823.01 & $-$ & $5536\pm79$ & $202.117\pm0.001$ & 0.667 & $5.8\pm0.53$ & $1.59\pm0.38$ & $0.96\pm0.05$ & $563\pm57$ & 13.92 \\
K01058.01 & $-$ & $3337\pm86$ & $5.67\pm0$ & 0.034 & $5.85\pm2.015$ & $3.22\pm2.55$ & $0.16\pm0.07$ & $32\pm12$ & 13.78 \\
K00683.01 & $-$ & $5799\pm110$ & $278.124\pm0$ & 0.842 & $5.86\pm0.72$ & $1.58\pm0.51$ & $1.03\pm0.07$ & $622\pm73$ & 13.71 \\
K05375.01 & $-$ & $5142\pm150$ & $285.375\pm0.004$ & 0.794 & $5.94\pm4.05$ & $7.56\pm11.19$ & $0.82\pm0.21$ & $1138\pm769$ & 13.86 \\
K05833.01 & $-$ & $6261\pm174$ & $440.171\pm0.006$ & 1.145 & $5.97\pm1.53$ & $2.97\pm1.85$ & $1.03\pm0.16$ & $809\pm200$ & 13.01 \\
K02076.02 & 1085b & $6063\pm181$ & $219.322\pm0.001$ & 0.739 & $6.11\pm1.085$ & $2.27\pm1.08$ & $1.12\pm0.14$ & $1314\pm270$ & 15.27 \\
K02681.01 & 397c & $5307\pm100$ & $135.499\pm0.001$ & 0.480 & $6.18\pm0.56$ & $1.83\pm0.47$ & $0.78\pm0.05$ & $983\pm76$ & 16.00 \\
K05416.01 & 1628b & $3869\pm140$ & $76.378\pm0.002$ & 0.295 & $6.28\pm0.6$ & $0.79\pm0.26$ & $0.59\pm0.06$ & $418\pm56$ & 16.60 \\
K01783.02 & $-$ & $5791\pm111$ & $284.063\pm0.002$ & 0.845 & $6.36\pm1.105$ & $2.52\pm1.07$ & $1\pm0.08$ & $913\pm157$ & 13.93 \\
K02689.01 & $-$ & $5594\pm186$ & $165.345\pm0$ & 0.547 & $6.98\pm1.175$ & $1.94\pm0.91$ & $0.8\pm0.08$ & $1001\pm191$ & 15.55 \\
K05278.01 & $-$ & $5330\pm187$ & $281.592\pm0.001$ & 0.776 & $7.22\pm0.885$ & $0.61\pm0.24$ & $0.8\pm0.08$ & $911\pm133$ & 15.87 \\
K03791.01 & 460b & $6340\pm190$ & $440.784\pm0.001$ & 1.146 & $7.23\pm2$ & $2.14\pm1.44$ & $1.03\pm0.15$ & $917\pm242$ & 13.77 \\
K01375.01 & $-$ & $6018\pm120$ & $321.212\pm0$ & 0.945 & $7.25\pm1.165$ & $2.18\pm0.87$ & $1.09\pm0.09$ & $755\pm129$ & 13.71 \\
K03263.01 & $-$ & $3638\pm76$ & $76.879\pm0$ & 0.275 & $7.71\pm0.83$ & $0.4\pm0.12$ & $0.47\pm0.05$ & $220\pm28$ & 15.95 \\
K01431.01 & $-$ & $5597\pm112$ & $345.159\pm0$ & 0.975 & $7.79\pm0.745$ & $0.8\pm0.22$ & $1.03\pm0.06$ & $456\pm48$ & 13.46 \\
K01439.01 & 849b & $5910\pm113$ & $394.625\pm0.001$ & 1.109 & $7.79\pm1.585$ & $2.66\pm1.28$ & $1.16\pm0.13$ & $740\pm147$ & 12.85 \\
K01411.01 & $-$ & $5716\pm109$ & $305.076\pm0$ & 0.912 & $7.82\pm1.045$ & $1.54\pm0.53$ & $1.08\pm0.07$ & $537\pm75$ & 13.38 \\
K00950.01 & $-$ & $3748\pm59$ & $31.202\pm0$ & 0.150 & $8.31\pm0.575$ & $1.59\pm0.32$ & $0.46\pm0.03$ & $237\pm21$ & 15.80 \\
K05071.01 & $-$ & $6032\pm211$ & $180.412\pm0.001$ & 0.637 & $8.86\pm1.73$ & $2.78\pm1.47$ & $1.06\pm0.14$ & $1373\pm301$ & 15.66 \\
K03663.01 & 86b & $5725\pm108$ & $282.525\pm0$ & 0.836 & $8.98\pm0.89$ & $1.15\pm0.31$ & $0.97\pm0.06$ & $328\pm35$ & 12.62 \\
K00620.03 & 51c & $6018\pm107$ & $85.312\pm0.003$ & 0.384 & $9\pm2.25$ & $7.05\pm8$ & $1.05\pm0.14$ & $927\pm205$ & 14.67 \\
K01477.01 & $-$ & $5270\pm79$ & $169.498\pm0.001$ & 0.575 & $9.06\pm0.59$ & $1.29\pm0.24$ & $0.9\pm0.05$ & $1053\pm78$ & 15.92 \\
K03678.01 & 1513b & $5650\pm186$ & $160.885\pm0$ & 0.542 & $9.09\pm2.53$ & $3.4\pm2.34$ & $0.82\pm0.09$ & $410\pm112$ & 12.89 \\
K08007.01 & $-$ & $3391\pm42$ & $67.177\pm0$ & 0.218 & $9.66\pm1.115$ & $0.24\pm0.07$ & $0.3\pm0.04$ & $135\pm18$ & 16.06 \\
K00620.02 & 51d & $6018\pm107$ & $130.194\pm0.004$ & 0.509 & $9.7\pm0.5$ & $4.01\pm4.56$ & $1.05\pm0.14$ & $927\pm205$ & 14.67 \\
K01681.04 & $-$ & $3638\pm80$ & $21.914\pm0$ & 0.117 & $10.39\pm1.26$ & $2.01\pm0.66$ & $0.45\pm0.05$ & $203\pm30$ & 15.86 \\
K00868.01 & $-$ & $4245\pm85$ & $235.999\pm0$ & 0.653 & $10.59\pm0.435$ & $0.29\pm0.05$ & $0.67\pm0.03$ & $358\pm22$ & 15.17 \\
K01466.01 & $-$ & $4810\pm76$ & $281.563\pm0$ & 0.766 & $10.83\pm0.535$ & $0.49\pm0.08$ & $0.76\pm0.04$ & $855\pm55$ & 15.96 \\
K00351.01 & 90h & $5970\pm119$ & $331.597\pm0$ & 0.965 & $10.89\pm1.61$ & $1.76\pm0.66$ & $1.09\pm0.08$ & $809\pm118$ & 13.80 \\
K00433.02 & 553c & $5234\pm103$ & $328.24\pm0$ & 0.908 & $10.99\pm0.77$ & $0.6\pm0.13$ & $0.93\pm0.05$ & $706\pm46$ & 14.92 \\
K05329.01 & $-$ & $6108\pm211$ & $200.235\pm0.001$ & 0.686 & $10.99\pm2.305$ & $2.64\pm1.47$ & $1.07\pm0.15$ & $1207\pm269$ & 15.39 \\
K03811.01 & $-$ & $5631\pm76$ & $290.14\pm0$ & 0.843 & $11.58\pm2.045$ & $2.02\pm0.82$ & $0.95\pm0.06$ & $738\pm130$ & 13.91 \\
K03801.01 & $-$ & $5672\pm76$ & $288.313\pm0.001$ & 0.846 & $13.21\pm2.185$ & $1.93\pm0.74$ & $0.97\pm0.07$ & $1837\pm318$ & 16.00 \\
K01268.01 & $-$ & $5798\pm78$ & $268.941\pm0.001$ & 0.827 & $13.57\pm2.305$ & $2.53\pm1$ & $1.04\pm0.08$ & $1262\pm219$ & 14.81 \\
\enddata
\tablenotetext{*}{Semi major axis}
\end{deluxetable*}


\floattable
\clearpage
\newpage
\startlongtable
\begin{deluxetable*}{llllllll}
  \tablewidth{0pc}
 \tabletypesize{\tiny}
  \tablecaption{Radial Velocity, Hill Radius \& Angular Separation Calculations for HZ Candidates with $R_p > 3 \ R_\oplus$.
\label{c4HRAS}}
\tablehead{
\colhead{KOI name}  & \colhead{Kepler} & \colhead{Planet Mass} & \colhead{Hill Radius} &  \colhead{$\alpha \arcsec Planet-Star$} & \colhead{$\alpha \arcsec Moon (HR)$} & \colhead{$\alpha \arcsec Moon  (\onethird HR) $} & \colhead{Radial Velocity} \\
\colhead{} & \colhead{} & \colhead{$M_\oplus$} & \colhead{AU}  & \colhead{$\mu$ arcsec} & \colhead{$\mu$ arcsec} & \colhead{$\mu$ arcsec} & \colhead{m/s}} 
\startdata
K03086.01 & $-$ & $6.44\pm0.98$ & $0.0114\pm0.0006$ & $570\pm48$ & $11.3\pm1.1$ & $3.78\pm0.37$ & $0.84\pm0.15$ \\
K06786.01 & $-$ & $6.44\pm2.44$ & $0.0216\pm0.0029$ & $361\pm62$ & $6.77\pm1.5$ & $2.26\pm0.49$ & $0.54\pm0.23$ \\
K02691.01 & $-$ & $6.62\pm1.12$ & $0.0078\pm0.0005$ & $834\pm93$ & $17.4\pm2.3$ & $5.81\pm0.75$ & $1.13\pm0.24$ \\
K01581.02 & 896b & $6.66\pm2.01$ & $0.0102\pm0.0011$ & $558\pm102$ & $11\pm2.4$ & $3.67\pm0.78$ & $0.89\pm0.29$ \\
K08156.01 & $-$ & $6.88\pm2.96$ & $0.019\pm0.0029$ & $1070\pm263$ & $19.4\pm5.6$ & $6.44\pm1.86$ & $0.56\pm0.27$ \\
K07700.01 & $-$ & $6.92\pm2.82$ & $0.0275\pm0.0039$ & $1870\pm414$ & $34.5\pm9.1$ & $11.5\pm3.03$ & $0.48\pm0.22$ \\
K04016.01 & 1540b & $6.95\pm0.54$ & $0.0094\pm0.0003$ & $1510\pm93$ & $32\pm2.2$ & $10.6\pm0.73$ & $1.09\pm0.11$ \\
K05706.01 & 1636b & $7.18\pm2.67$ & $0.0214\pm0.0028$ & $727\pm159$ & $13.5\pm3.4$ & $4.47\pm1.14$ & $0.56\pm0.23$ \\
K02210.02 & 1143c & $7.3\pm0.66$ & $0.0134\pm0.0005$ & $1070\pm67$ & $22.1\pm1.6$ & $7.42\pm0.54$ & $0.9\pm0.1$ \\
K08276.01 & $-$ & $7.3\pm3.1$ & $0.0201\pm0.003$ & $1170\pm268$ & $21.3\pm5.8$ & $7.1\pm1.94$ & $0.56\pm0.26$ \\
K04121.01 & 1554b & $7.33\pm1.59$ & $0.0129\pm0.001$ & $543\pm67$ & $11.1\pm1.6$ & $3.69\pm0.54$ & $0.89\pm0.2$ \\
K05622.01 & 1635b & $7.33\pm2.03$ & $0.0229\pm0.0023$ & $1180\pm201$ & $24.3\pm4.8$ & $8.05\pm1.59$ & $0.67\pm0.21$ \\
K07982.01 & $-$ & $7.41\pm2.94$ & $0.0199\pm0.0028$ & $716\pm166$ & $13.9\pm3.8$ & $4.6\pm1.25$ & $0.65\pm0.28$ \\
K03946.01 & 1533b & $7.49\pm2.51$ & $0.0175\pm0.002$ & $1310\pm212$ & $23.8\pm4.7$ & $7.9\pm1.57$ & $0.61\pm0.22$ \\
K08232.01 & $-$ & $7.6\pm3.45$ & $0.0127\pm0.002$ & $706\pm173$ & $14.7\pm4.3$ & $4.86\pm1.41$ & $0.95\pm0.46$ \\
K05625.01 & $-$ & $7.68\pm1.69$ & $0.0092\pm0.0007$ & $463\pm69$ & $10.3\pm1.7$ & $3.47\pm0.58$ & $1.28\pm0.34$ \\
K02073.01 & 357d & $8.08\pm9.36$ & $0.0053\pm0.0021$ & $319\pm21$ & $6.87\pm2.8$ & $2.33\pm0.94$ & $1.64\pm1.91$ \\
K02686.01 & $-$ & $8.08\pm0.78$ & $0.0139\pm0.0005$ & $2350\pm150$ & $52.1\pm3.8$ & $17.2\pm1.26$ & $1.06\pm0.13$ \\
K01855.01 & $-$ & $8.16\pm1.38$ & $0.0059\pm0.0004$ & $832\pm92$ & $19.8\pm2.6$ & $6.71\pm0.87$ & $1.9\pm0.41$ \\
K02828.02 & $-$ & $8.2\pm1.45$ & $0.025\pm0.0016$ & $1500\pm185$ & $32.5\pm4.5$ & $10.8\pm1.5$ & $0.76\pm0.15$ \\
K02926.05 & $-$ & $8.24\pm0.88$ & $0.0071\pm0.0003$ & $698\pm58$ & $16.7\pm1.6$ & $5.65\pm0.52$ & $1.74\pm0.22$ \\
K08286.01 & $-$ & $8.52\pm2.81$ & $0.0133\pm0.0015$ & $383\pm78$ & $8.04\pm1.9$ & $2.66\pm0.62$ & $0.99\pm0.35$ \\
K01830.02 & 967c & $8.6\pm1.01$ & $0.0137\pm0.0006$ & $1250\pm92$ & $27.3\pm2.3$ & $9.17\pm0.79$ & $1.07\pm0.15$ \\
K00951.02 & 258c & $8.81\pm11.55$ & $0.0042\pm0.0019$ & $125\pm35$ & $2.72\pm1.5$ & $0.91\pm0.48$ & $1.98\pm2.6$ \\
K01986.01 & 1038b & $8.81\pm0.97$ & $0.0113\pm0.0005$ & $864\pm60$ & $18.6\pm1.5$ & $6.27\pm0.52$ & $1.17\pm0.15$ \\
K01527.01 & $-$ & $8.93\pm1.53$ & $0.0136\pm0.0008$ & $837\pm80$ & $18.3\pm2.1$ & $6.06\pm0.68$ & $1.09\pm0.21$ \\
K05790.01 & $-$ & $9.23\pm1.02$ & $0.0128\pm0.0005$ & $888\pm61$ & $19.9\pm1.6$ & $6.69\pm0.53$ & $1.2\pm0.16$ \\
K08193.01 & $-$ & $9.27\pm2.91$ & $0.0211\pm0.0023$ & $892\pm162$ & $18.9\pm4$ & $6.27\pm1.33$ & $0.84\pm0.29$ \\
K08275.01 & $-$ & $9.44\pm2.25$ & $0.0221\pm0.0019$ & $1030\pm160$ & $22.7\pm4$ & $7.59\pm1.35$ & $0.9\pm0.24$ \\
K01070.02 & 266c & $10\pm9.46$ & $0.01\pm0.0032$ & $293\pm53$ & $6.4\pm2.4$ & $2.11\pm0.78$ & $1.39\pm1.32$ \\
K07847.01 & $-$ & $10.17\pm6.18$ & $0.023\pm0.0048$ & $503\pm164$ & $10.5\pm4.1$ & $3.52\pm1.36$ & $0.82\pm0.53$ \\
K00401.02 & 149d & $10.3\pm3.45$ & $0.0127\pm0.0014$ & $1060\pm109$ & $23.5\pm3.6$ & $7.76\pm1.17$ & $1.27\pm0.43$ \\
K01707.02 & 315c & $11.16\pm5.03$ & $0.0185\pm0.0028$ & $731\pm99$ & $17.1\pm3.5$ & $5.73\pm1.16$ & $1.21\pm0.56$ \\
K05581.01 & 1634b & $11.71\pm6.01$ & $0.0231\pm0.0041$ & $1030\pm276$ & $22.7\pm7.3$ & $7.55\pm2.42$ & $0.97\pm0.52$ \\
K01258.03 & $-$ & $11.85\pm4.03$ & $0.0125\pm0.0015$ & $448\pm90$ & $10.3\pm2.4$ & $3.45\pm0.81$ & $1.45\pm0.54$ \\
K02683.01 & $-$ & $12.75\pm3.51$ & $0.0115\pm0.0011$ & $499\pm78$ & $12.1\pm2.2$ & $4.01\pm0.73$ & $1.76\pm0.55$ \\
K00881.02 & 712c & $12.94\pm1.45$ & $0.0171\pm0.0007$ & $788\pm55$ & $20\pm1.6$ & $6.67\pm0.54$ & $1.59\pm0.22$ \\
K01429.01 & $-$ & $13.68\pm2.85$ & $0.0163\pm0.0012$ & $551\pm60$ & $13.2\pm1.8$ & $4.38\pm0.58$ & $1.5\pm0.34$ \\
K00902.01 & $-$ & $14.18\pm2.34$ & $0.0091\pm0.0006$ & $872\pm108$ & $26.2\pm3.7$ & $8.63\pm1.21$ & $3.18\pm0.63$ \\
K05929.01 & $-$ & $14.89\pm5.16$ & $0.029\pm0.0035$ & $1490\pm322$ & $37.2\pm9.2$ & $12.4\pm3.07$ & $1.25\pm0.48$ \\
K00179.02 & 458b & $19.68\pm5.98$ & $0.0365\pm0.0038$ & $1560\pm241$ & $40.4\pm7.5$ & $13.5\pm2.52$ & $1.4\pm0.45$ \\
K03823.01 & $-$ & $19.68\pm3.5$ & $0.0182\pm0.0011$ & $1180\pm120$ & $32.3\pm3.8$ & $10.8\pm1.28$ & $2.2\pm0.43$ \\
K01058.01 & $-$ & $19.96\pm13.39$ & $0.0017\pm0.0004$ & $1070\pm407$ & $53.7\pm23.9$ & $18.9\pm8.45$ & $23.89\pm21.28$ \\
K00683.01 & $-$ & $20.02\pm4.79$ & $0.0227\pm0.0019$ & $1350\pm159$ & $36.5\pm5.3$ & $12.2\pm1.76$ & $1.92\pm0.5$ \\
K05375.01 & $-$ & $20.49\pm27.21$ & $0.0232\pm0.0105$ & $697\pm471$ & $20.4\pm16.6$ & $6.76\pm5.5$ & $2.28\pm3.14$ \\
K05833.01 & $-$ & $20.66\pm10.32$ & $0.0311\pm0.0054$ & $1420\pm350$ & $38.4\pm11.6$ & $12.9\pm3.88$ & $1.7\pm0.93$ \\
K02076.02 & 1085b & $21.49\pm7.43$ & $0.0198\pm0.0024$ & $562\pm116$ & $15.1\pm3.6$ & $5.02\pm1.2$ & $2.12\pm0.82$ \\
K02681.01 & 397c & $21.91\pm3.87$ & $0.0146\pm0.0009$ & $488\pm38$ & $14.8\pm1.5$ & $4.98\pm0.49$ & $3.21\pm0.63$ \\
K05416.01 & 1628b & $22.51\pm4.19$ & $0.01\pm0.0007$ & $706\pm95$ & $23.9\pm3.6$ & $7.89\pm1.19$ & $4.84\pm1.11$ \\
K01783.02 & $-$ & $23\pm7.78$ & $0.0241\pm0.0028$ & $925\pm159$ & $26.4\pm5.5$ & $8.76\pm1.82$ & $2.24\pm0.8$ \\
K02689.01 & $-$ & $26.93\pm8.83$ & $0.0177\pm0.002$ & $546\pm104$ & $17.7\pm3.9$ & $5.89\pm1.31$ & $3.65\pm1.31$ \\
K05278.01 & $-$ & $28.52\pm6.81$ & $0.0256\pm0.0022$ & $852\pm124$ & $28.1\pm4.8$ & $9.33\pm1.58$ & $3.24\pm0.91$ \\
K03791.01 & 460b & $28.59\pm15.4$ & $0.0347\pm0.0064$ & $1250\pm329$ & $37.8\pm12.2$ & $12.6\pm4.07$ & $2.35\pm1.36$ \\
K01375.01 & $-$ & $28.72\pm8.99$ & $0.0281\pm0.003$ & $1250\pm214$ & $37.2\pm7.5$ & $12.4\pm2.51$ & $2.53\pm0.85$ \\
K03263.01 & $-$ & $31.88\pm6.68$ & $0.0112\pm0.0009$ & $1250\pm159$ & $50.8\pm7.7$ & $16.8\pm2.53$ & $7.96\pm2.02$ \\
K01431.01 & $-$ & $32.44\pm6.04$ & $0.0308\pm0.002$ & $2140\pm225$ & $67.6\pm8.4$ & $22.6\pm2.8$ & $2.9\pm0.58$ \\
K01439.01 & 849b & $32.44\pm12.86$ & $0.0336\pm0.0046$ & $1500\pm298$ & $45.4\pm11$ & $15.1\pm3.66$ & $2.56\pm1.09$ \\
K01411.01 & $-$ & $32.65\pm8.5$ & $0.0284\pm0.0025$ & $1700\pm237$ & $52.9\pm8.7$ & $17.7\pm2.92$ & $2.94\pm0.81$ \\
K00950.01 & $-$ & $36.19\pm4.88$ & $0.0064\pm0.0003$ & $633\pm56$ & $27\pm2.7$ & $8.87\pm0.89$ & $12.32\pm2.01$ \\
K05071.01 & $-$ & $40.35\pm15.35$ & $0.0215\pm0.0029$ & $464\pm102$ & $15.7\pm4$ & $5.25\pm1.35$ & $4.41\pm1.87$ \\
K03663.01 & 86b & $41.28\pm7.97$ & $0.0292\pm0.002$ & $2550\pm272$ & $89\pm11.3$ & $29.6\pm3.75$ & $4.09\pm0.88$ \\
K00620.03 & 51c & $41.43\pm20.18$ & $0.0131\pm0.0022$ & $414\pm92$ & $14.1\pm3.9$ & $4.75\pm1.32$ & $5.81\pm3.02$ \\
K01477.01 & $-$ & $41.9\pm5.32$ & $0.0207\pm0.0009$ & $546\pm41$ & $19.7\pm1.7$ & $6.55\pm0.56$ & $5.19\pm0.76$ \\
K03678.01 & 1513b & $42.14\pm22.84$ & $0.0202\pm0.0037$ & $1320\pm361$ & $49.3\pm16.2$ & $16.3\pm5.38$ & $5.66\pm3.2$ \\
K08007.01 & $-$ & $46.71\pm10.5$ & $0.0117\pm0.001$ & $1610\pm214$ & $86.5\pm13.7$ & $28.8\pm4.56$ & $16.25\pm4.89$ \\
K00620.02 & 51d & $47.04\pm4.72$ & $0.0181\pm0.001$ & $549\pm121$ & $19.5\pm4.5$ & $6.47\pm1.48$ & $5.73\pm1.19$ \\
K01681.04 & $-$ & $52.85\pm12.48$ & $0.0058\pm0.0005$ & $578\pm87$ & $28.6\pm4.9$ & $9.36\pm1.62$ & $20.56\pm5.87$ \\
K00868.01 & $-$ & $54.59\pm4.37$ & $0.0284\pm0.0009$ & $1830\pm112$ & $79.4\pm5.5$ & $26.6\pm1.84$ & $7.41\pm0.77$ \\
K01466.01 & $-$ & $56.7\pm5.46$ & $0.0323\pm0.0012$ & $896\pm58$ & $37.8\pm2.8$ & $12.6\pm0.94$ & $6.67\pm0.78$ \\
K00351.01 & 90h & $57.23\pm16.48$ & $0.0362\pm0.0036$ & $1190\pm174$ & $44.8\pm7.9$ & $15\pm2.64$ & $4.99\pm1.54$ \\
K00433.02 & 553c & $58.13\pm7.93$ & $0.0361\pm0.0017$ & $1290\pm84$ & $51.2\pm4.1$ & $17\pm1.37$ & $5.67\pm0.87$ \\
K05329.01 & $-$ & $58.13\pm23.75$ & $0.026\pm0.0037$ & $568\pm127$ & $21.5\pm5.7$ & $7.21\pm1.91$ & $6.06\pm2.74$ \\
K03811.01 & $-$ & $63.52\pm21.85$ & $0.0343\pm0.004$ & $1140\pm201$ & $46.4\pm9.8$ & $15.4\pm3.26$ & $6.36\pm2.27$ \\
K03801.01 & $-$ & $79.4\pm25.58$ & $0.0368\pm0.004$ & $461\pm80$ & $20\pm4.1$ & $6.7\pm1.37$ & $7.85\pm2.65$ \\
K01268.01 & $-$ & $83.1\pm27.5$ & $0.0356\pm0.004$ & $655\pm114$ & $28.2\pm5.8$ & $9.43\pm1.95$ & $8.01\pm2.77$ \\
\enddata
\end{deluxetable*}


\begin{deluxetable*}{llllllll}
\tablewidth{0pc}
\tabletypesize{\scriptsize}
\tablecaption{Radial Velocity Semi-amplitude calculations for Category 4 HZ candidates with $R_p > 10 \ R_\oplus$.
\label{c4RVSA}}
\tablehead{
\colhead{KOI name} & \colhead{Kepler} & \colhead{Period} & \colhead{Planet Radius} & 
\colhead{Stellar Mass} & \colhead{RV ($M_{sat}$)} & \colhead{RV ($M_J$)} & \colhead{RV ($13 M_J$)} \\
\colhead{} & \colhead{} & \colhead{Days} & \colhead{$R_\oplus$}  & \colhead{$M_\star$} & \colhead{m/s} & \colhead{m/s} & \colhead{m/s} 
}
\startdata
K01681.04 &  & $21.914\pm0.0002$ & $10.39\pm1.26$ & $0.45\pm0.051$ & $37.03\pm5.94$ & $123.73\pm20.08$ & $1621.95\pm258.66$ \\
K00868.01 &  & $235.999\pm0.0003$ & $10.59\pm0.435$ & $0.666\pm0.031$ & $12.91\pm0.86$ & $43.13\pm3.06$ & $563.9\pm38.53$ \\
K01466.01 &  & $281.563\pm0.0004$ & $10.83\pm0.535$ & $0.755\pm0.036$ & $11.2\pm0.76$ & $37.4\pm2.71$ & $488.67\pm34.16$ \\
K00351.01 & 90h & $331.597\pm0.0003$ & $10.89\pm1.61$ & $1.089\pm0.084$ & $8.3\pm0.91$ & $27.74\pm3.11$ & $361.88\pm39.94$ \\
K00433.02 & 553c & $328.24\pm0.0004$ & $10.99\pm0.77$ & $0.927\pm0.045$ & $9.28\pm0.64$ & $30.99\pm2.28$ & $404.54\pm28.79$ \\
K05329.01 &  & $200.235\pm0.0006$ & $10.99\pm2.305$ & $1.072\pm0.146$ & $9.93\pm1.91$ & $33.17\pm6.45$ & $432.68\pm83.35$ \\
K03811.01 &  & $290.14\pm0.0003$ & $11.58\pm2.045$ & $0.947\pm0.064$ & $9.53\pm0.91$ & $31.84\pm3.16$ & $415.53\pm40.36$ \\
K03801.01 &  & $288.313\pm0.0005$ & $13.21\pm2.185$ & $0.969\pm0.068$ & $9.41\pm0.94$ & $31.42\pm3.23$ & $410.03\pm41.29$ \\
K01268.01 &  & $268.941\pm0.0005$ & $13.57\pm2.305$ & $1.041\pm0.075$ & $9.18\pm0.94$ & $30.65\pm3.23$ & $399.95\pm41.32$ \\
\enddata
\end{deluxetable*}

\begin{deluxetable*}{lllllllll}
\tablewidth{0pc}
\tabletypesize{\tiny}
\tablecaption{Hill Radii calculations for Category 4 HZ candidates with $R_p > 10 \ R_\oplus$.
\label{c4HR}}
\tablehead{
\colhead{KOI name} & \colhead{Kepler} & \colhead{Planet Radius} &  \colhead{Hill Radius ($M_{sat}$)} & \colhead{Hill Radius ($M_J$)} & \colhead{Hill Radius (13 $M_J$)} & \colhead{$\alpha \arcsec$ ($M_{sat}$) \tablenotemark{a}} & 
\colhead{$\alpha \arcsec$ ($M_J$) \tablenotemark{b} } & \colhead{$\alpha \arcsec$ ($13M_J$) \tablenotemark{c} } \\
\colhead{} & \colhead{} & \colhead{$R_\oplus$} & \colhead{AU}  & \colhead{AU} & \colhead{AU} & \colhead{$ \mu$ arcsec} & \colhead{$ \mu$ arcsec} & \colhead{$ \mu$ arcsec}
}
\startdata
K01681.04 &  & $10.39\pm1.26$ & $0.007\pm0.0003$ & $0.0105\pm0.0004$ & $0.0246\pm0.0009$ & $28.6\pm4.9$ & $9.4\pm1.6$ & $578\pm87$ \\
K00868.01 &  & $10.59\pm0.435$ & $0.0342\pm0.0005$ & $0.0511\pm0.0009$ & $0.1201\pm0.002$ & $79.4\pm5.5$ & $26.6\pm1.8$ & $1830\pm112$ \\
K01466.01 &  & $10.83\pm0.535$ & $0.0384\pm0.0006$ & $0.0574\pm0.001$ & $0.135\pm0.0023$ & $37.8\pm2.8$ & $12.6\pm0.9$ & $896\pm58$ \\
K00351.01 & 90h & $10.89\pm1.61$ & $0.0429\pm0.0011$ & $0.0641\pm0.0017$ & $0.1506\pm0.004$ & $44.8\pm7.9$ & $15\pm2.6$ & $1190\pm174$ \\
K00433.02 & 553c & $10.99\pm0.77$ & $0.0425\pm0.0007$ & $0.0636\pm0.0012$ & $0.1495\pm0.0026$ & $51.2\pm4.1$ & $17\pm1.4$ & $1290\pm84$ \\
K05329.01 &  & $10.99\pm2.305$ & $0.0306\pm0.0014$ & $0.0458\pm0.0021$ & $0.1076\pm0.0049$ & $21.5\pm5.7$ & $7.2\pm1.9$ & $568\pm127$ \\
K03811.01 &  & $11.58\pm2.045$ & $0.0392\pm0.0009$ & $0.0586\pm0.0014$ & $0.1378\pm0.0032$ & $46.4\pm9.8$ & $15.4\pm3.3$ & $1140\pm201$ \\
K03801.01 &  & $13.21\pm2.185$ & $0.039\pm0.0009$ & $0.0584\pm0.0015$ & $0.1372\pm0.0033$ & $20\pm4.1$ & $6.7\pm1.4$ & $461\pm80$ \\
K01268.01 &  & $13.57\pm2.305$ & $0.0373\pm0.0009$ & $0.0557\pm0.0014$ & $0.131\pm0.0033$ & $28.2\pm5.8$ & $9.4\pm2$ & $655\pm114$\\
\enddata
\tablenotetext{a}{Angular separation of exomoon at full Hill radius for $M_p = M_\mathrm{sat}$.}
\tablenotetext{b}{Angular separation of exomoon at full Hill radius for $M_p = M_J$.}
\tablenotetext{c}{Angular separation of exomoon at full Hill radius for $M_p = 13 M_J$.}
\end{deluxetable*}


We plot a histogram of the effective temperatures of Kepler host stars
to determine if there is a similar distribution of temperatures among
both the HZ candidates and the full catalog. 
\begin{figure}
\begin{center}
 \includegraphics[width=\columnwidth]{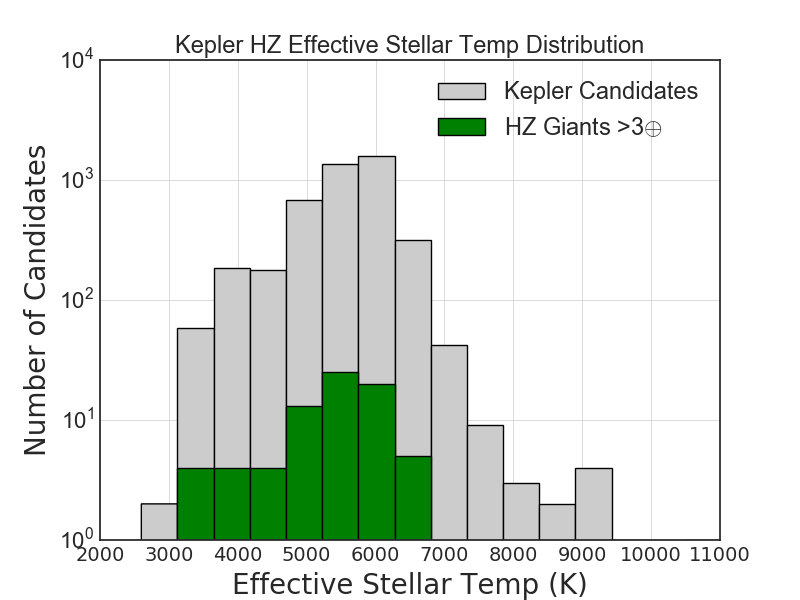}
  \caption{Stellar temperature distributions. Habitable zone Kepler candidates in green overlays the distribution of the full Kepler catalog in gray. The histograms show that there is a similar distribution of temperatures among both the HZ candidates and the full Kepler catalog. While the distribution of the habitable zone candidates drops off at 7000K, this could be a false upper limit as the habitable zone of stars with greater effective temperature lies further away from the star and current transit detection methods are less sensitive to planets at these longer orbits.}
   \label{fig:Stell}
  \end{center}
\end{figure}

Figure \ref{fig:Stell} shows the stellar temperature distributions for
both the HZ Kepler candidates (green) as well as the full Kepler
catalog (gray). The histograms show that there is a similar
distribution of temperatures among both the HZ candidates and the full
catalog, with the HZ host star temperatures dropping off (around)
7000K. As the habitable zone of stars with greater effective temperatures will lie further away from the star, planets in this zone are harder to detect. Thus this drop is likely a false upper limit.

Using the calculations from our Tables above, we plot the Kepler
magnitude of the host star of both the unconfirmed and confirmed HZ
planets and their expected radial velocity signatures to determine the
expected detectability of these planets.

\begin{figure}
\begin{center}
  \includegraphics[width=\columnwidth]{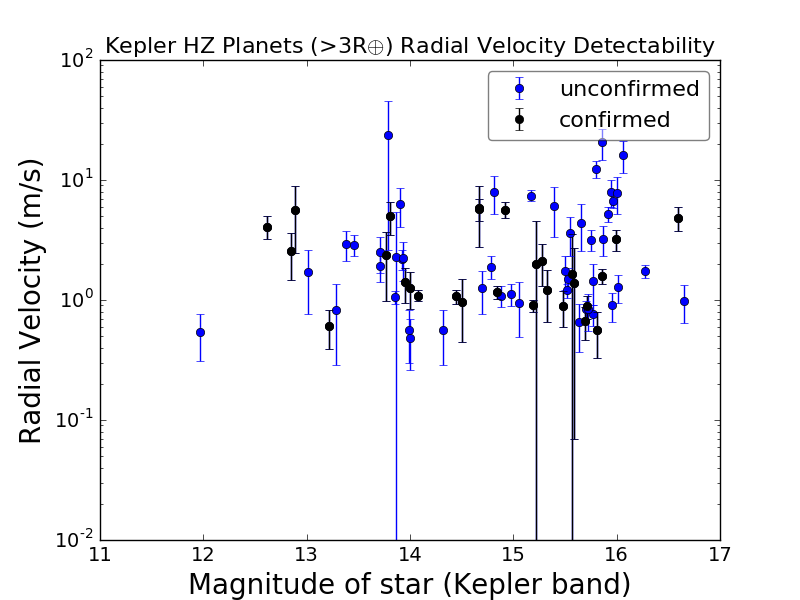}
  \caption{We plot the Kepler magnitude of the host star of both the unconfirmed and confirmed HZ planets and their expected radial velocity signatures to determine the expected detectability of these planets. We find that a large majority of the planets in our list have an estimated radial velocity semi amplitude between 1 and 10 m/s. As the {\it Kepler} telescope was focused on a field faint stars, the planets listed are at the limit of the capabilities of current RV detection instruments. Future radial velocity missions to follow up on these candidates should focus on those found closest to the top left corner of the graph, where the brightest stars host candidates with large RV semi amplitudes.}
  \label{fig:RVMag1}
\end{center}
\end{figure}

Figure \ref{fig:RVMag1} shows the Kepler magnitude of the host
star of both the unconfirmed and confirmed HZ planets and their
expected radial velocity signatures.

\begin{figure}
\begin{center}
  \includegraphics[width=\columnwidth]{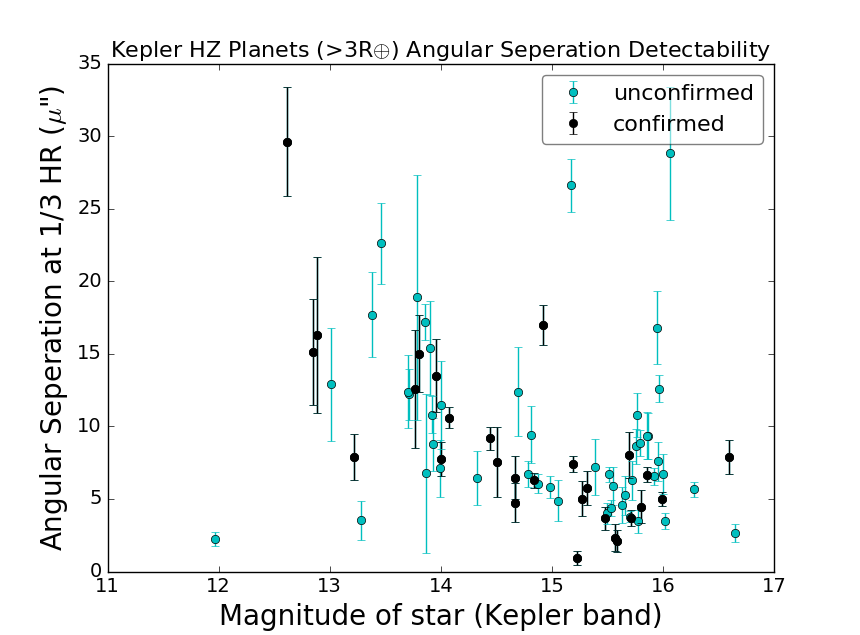}
  \caption{We plot the Kepler magnitude of the host star of both the unconfirmed and confirmed HZ
planets and their expected Angular separation to determine the expected detectability of these planets. Confirmed candidates are noted by black dots and unconfirmed candidates by teal dots. Note the Y axis is the angular separation at $\onethird$ Hill radius which we have taken as the typical distance of a stable moon. Future imaging missions will need the capabilities to resolve a separation between 1 – 35 $\mu$ arc seconds.}
  \label{fig:ASMag2}
\end{center}
\end{figure}

We then provide a similar plot in Figure \ref{fig:ASMag2}, this time plotting the Kepler magnitude of the host star of both the unconfirmed and confirmed HZ planets and their expected angular separations of a moon at the full Hill radius of the host planet.

\begin{figure}
\begin{center}
  \includegraphics[width=\columnwidth]{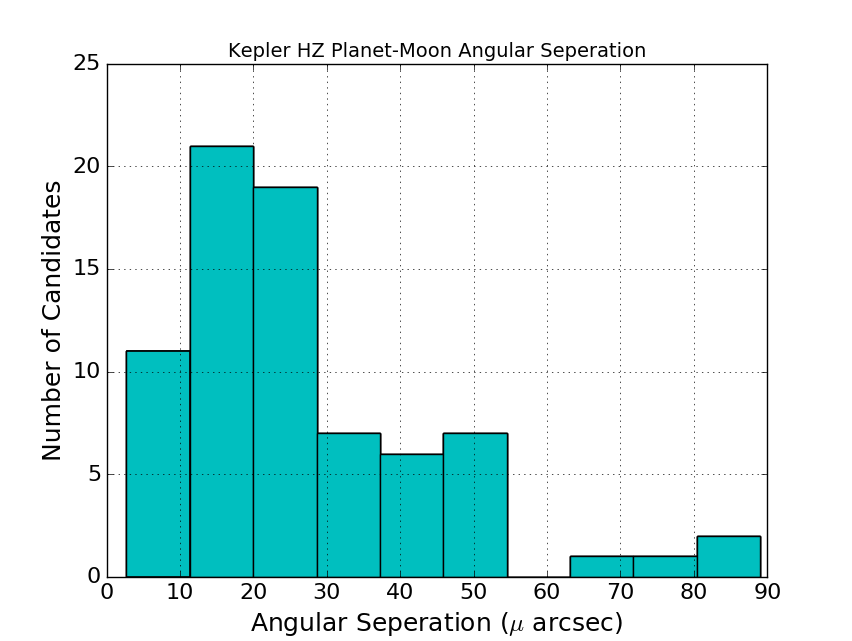}
  \caption{Here we show the distribution of Kepler habitable zone planets  ($ > 3R\oplus $) Planet - Moon angular separation, with moons positioned at the full Hill radii. Potential moons of giant planets found in the habitable zone will likely have a maximum angular separation from their host planet between 1 - 90 $\mu$ arc seconds. This information can be used for planning of imaging future missions if we assume Kepler candidates are representative of the entire population of stars and planets.}
  \label{fig:AS5}
\end{center}
\end{figure}

Figure \ref{fig:AS5} shows the distribution of the estimated planet - moon angular separation at the full Hill radii of the candidate. It can be seen that the resolution required to image a moon is between 1 - 90 $\mu $ arcseconds with the moon positioned at its maximum stable distance from the planet. If a potential moon resides within $\onethird$ Hill radius from the planet as expected, the resolution will need to improve as much again. Note these graphs do not take into account the separate calculations of angular separation for those planets $\geq 10R_\oplus$. 

\begin{figure}
\begin{center}
  \includegraphics[width=\columnwidth]{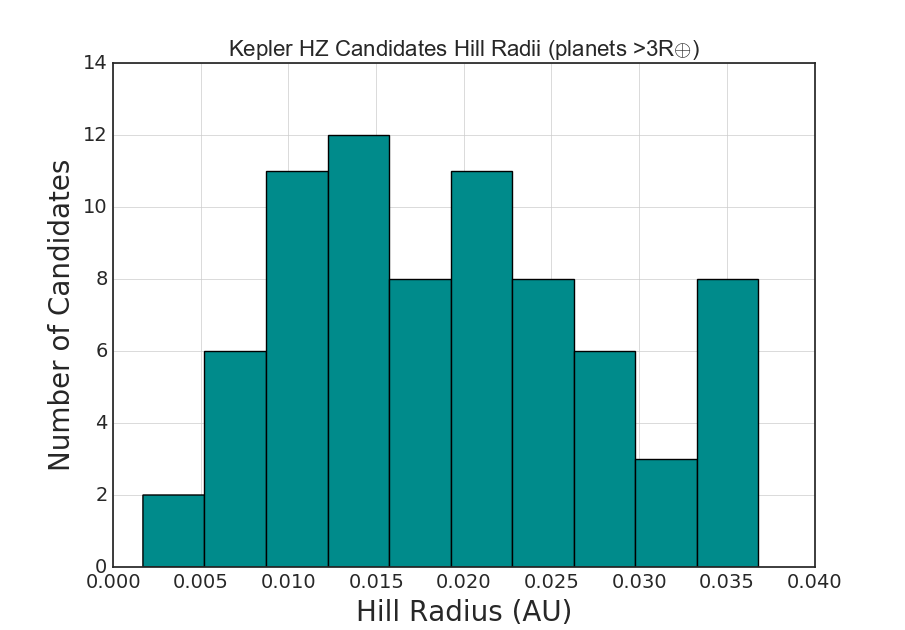}
  \caption{Here we show the distribution of Kepler habitable zone planets  ($ > 3R\oplus $) Hill radii. Potential moons of giant planets found in the habitable zone will likely have a maximum radius of gravitational influence between 5 - 35 milli AU. This information can be used for planning of imaging future missions as the Kepler candidates can be considered representative of the entire population of stars.}
  \label{fig:HR4}
\end{center}
\end{figure}

Figure \ref{fig:HR4} shows the distribution of the Hill radii of Kepler habitable zone planets  $ > 3R\oplus $. Potential moons of giant planets found in the habitable zone will likely have a maximum radius of gravitational influence between 5 - 35 Milli AU. If we assume a similar distribution exists around the entire population of giant planets found in the HZ, we can use this information to calculate the expected angular separation of a moon around the closest giant HZ planets. This can then be used for planning of future imaging missions. 

\begin{figure}
\begin{center}
  \includegraphics[width=\linewidth]{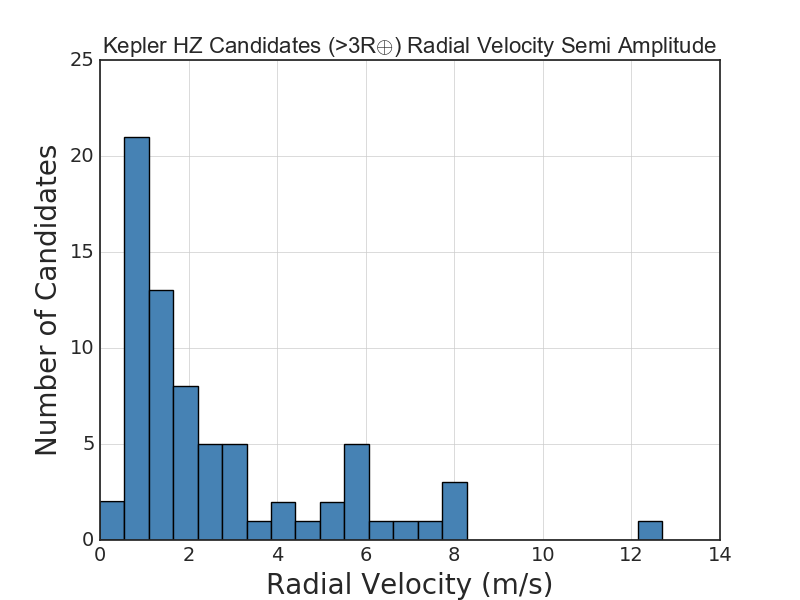}
  \caption{Here we show the distribution of Kepler habitable zone candidates  ($ > 3R\oplus $) estimated radial velocity semi amplitudes. As the giant planets we are investigating reside in the habitable zone of their star, the increased distance from the host star produces a relatively small RV semi amplitude, thus the majority of the candidates have estimated radial velocity semi amplitudes of \textless 2 m/s.}
  \label{fig:RV5}
\end{center}
\end{figure}

Finally, Figure \ref{fig:RV5} shows the distribution of the radial velocity semi amplitude of
the HZ candidates. While we estimate the majority of candidates will have a signature \textless 2 m/s, 
there are a number of planets that are likely to have significantly larger signatures and thus more easily detectable. However, as the Kepler stars are faint, even the largest of these signatures are on the limit of our current detection capabilities and so these planets will still be difficult to observe. Note this graph does not take into account the separate calculations of the radial 
velocity semi amplitude for those planets $\geq 10R_\oplus$.


\section{Discussion and Conclusions}
\label{Xmoon}

From our calculations in Section~\ref{hzgiant} we found the frequency
of giant planets ($R_p =$~3.0--25~$R_\oplus$) in the OHZ is $(6.5 \pm 1.9)\%$ for G stars, $(11.5 \pm 3.1)\%$ for K stars,
and $(6 \pm 6)\%$ for M stars. For comparison, the estimates of occurrence rates of terrestrial planets in the HZ for G-dwarf stars range from 2\% \citep{Fore14} to 22\%
\citep{Peti13} for GK dwarfs, but systematic errors dominate \citep{Burke15}. For M-dwarfs, the occurrence rates of terrestrial planets in the HZ is $\sim$20\% \citep{Dress15}. Therefore, it appears that the occurrence of large terrestrial moons orbiting giant planets in the HZ is less than the occurrence of terrestrial planets in the HZ. However this assumes that each giant planet is harboring only one large terrestrial exomoon. If giant planets can host multiple exomoons then the occurrence rates of moons would be comparable to that of terrestrial planets in the HZ of their star, and could potentially exceed them.


The calculations in Tables \ref{c4HRAS}, \ref{c4RVSA} and \ref{c4HR}
are intended for the design and observing
strategies of future RV surveys and direct imaging missions. We found that a large majority of the planets in our list have an estimated RV semi-amplitude between 1 and 10 m/s. While currently 1 m/s RV detection is regularly achieved around bright stars, the {\it Kepler} telescope was focused on a field faint stars, thus the planets included in our tables are at the limit of the capabilities of current RV detection. Precision RV capability is planned for the forthcoming generation of extremely large telescopes, such as the GMT-Consortium Large Earth Finder (G-CLEF) designed for the Giant Magellan Telescope (GMT) \citep{sze16}, further increasing the capabilities towards the measurement of masses for giant planets in the HZ. Future RV surveys to follow up these candidates should focus on those candidates with the largest estimated RV semi-amplitudes orbiting the brightest stars. 

Tidally heated exomoons can potentially be detected in direct imaging, if the contrast ratio of the satellite and the planet is favorable \citep{Pete13}. This is particularly beneficial for low mass stars, where the low stellar luminosity may aid in the detection of a tidally heated exomoon. However, the small inner working angle for low-mass stars will be unfavorable for characterization purposes.

A new approach was proposed for detection and characterization of exomoons based on spectroastrometry \citep{Agol15}. This method is based on the principle that the moon outshines the planet at certain wavelengths, and the centroid offset of the PSF (after suppressing the starlight with either a coronagraph or a starshade) observed in different wavelengths will enable one to detect an exomoon. For instance, the Moon outshines Earth at $\sim$2.7~$\mu$m. Ground-based facilities can possibly probe the HZs around M-dwarfs for exomoons, but large space-based telescopes, such as the 15m class LUVOIR, are necessary for obtaining sharper PSF and resolving the brightness.

If imaging of an exomoon orbiting a Kepler giant planet in the habitable zone is desired, instruments must have the capability to resolve a separation between $\thicksim~1~-~90~\mu$~arcseconds. The large distance and low apparent brightness of the Kepler stars makes them unideal for direct imaging. But if we assume the distribution of Hill radii (Figure \ref{fig:HR4}) calculated to surround the Kepler giant HZ planets to be representative of the larger giant HZ planet population, then our closest giant HZ planets could have exomoons with angular separations as large as $\thicksim~1~-~35~m$~arcseconds (assuming the closest giant HZ planets to reside between 1-10pc away).

Additional potential for exomoon detection lies in the method of microlensing, and has been demonstrated to be feasible with current survey capabilities for a subset of microlensing events \citep{lie10}. Furthermore, the microlensing detection technique is optimized for star--planet separations that are close to the snow line of the host stars \citep{gou10}, and simulations of stellar population distributions have shown that lens stars will predominately lie close to the near-side of the galactic center \citep{kan06}. A candidate microlensing exomoon was detected by \citet{ben14}, suggested to be a free-floating exoplanet-exomoon system. However, issues remain concerning the determination of the primary lens mass and any follow-up observations that would allow validation and characterization of such exomoon systems.

There is great habitability potential for the moons of giant exoplanets residing in
their HZ. These potentially terrestrial giant satellites could be the perfect hosts for 
life to form and take hold. Thermal and reflected radiation from the host planet and tidal effects increase the outer range of the HZ, creating a wider temperate zone in which a stable body 
may exist. There are, however, some caveats including the idea that giant planets in the HZ of their star may have migrated there \citep{Lunine01, Darriba17}. The moon of a giant planet migrating through the HZ may only have a short period in which the moon is considered habitable. Also, a planet that migrates inwards will eventually lose its moon(s) due to the shrinking Hill sphere of the planet \citep{Spal16}. Thus any giant planet that is in the HZ but still migrating inwards can quickly lose its moon as it moves closer to the host star.

\citep{Sart99} uncovered another factor potentially hindering the detection of these HZ moons when they found that multiple moons around a single planet may wash out any transit timing signal. And the small radius combined with the low contrast between planet and moon brightness mean transits are also unlikely to be a good method for detection. 

The occurrence rates calculated in Section 3 indicate a modest number of giant planets residing in the habitable zone of their star. Once imaging capabilities have improved, the detection of potentially habitable moons around these giant hosts should be more accessible. Until then we must continue to refine the properties of the giant host planets, starting with the radial velocity follow-up observations of the giant HZ candidates from our list.

\section*{Acknowledgements}  

This research has made use of the NASA Exoplanet Archive and the ExoFOP site, which are           
operated by the California Institute of Technology, under contract              
with the National Aeronautics and Space Administration under the                
Exoplanet Exploration Program. This work has also made use of the               
Habitable Zone Gallery at hzgallery.org. The results reported herein            
benefited from collaborations and/or information exchange within                
NASA's Nexus for Exoplanet System Science (NExSS) research                      
coordination network sponsored by NASA's Science Mission Directorate. 
The research shown here acknowledges use of the Hypatia Catalog Database, an online compilation of stellar abundance data as described in {Hinkel14}, which was supported by NASA's Nexus for Exoplanet System Science (NExSS) research coordination network and the Vanderbilt Initiative in Data-Intensive Astrophysics (VIDA).
This research has also made use of the VizieR catalogue access tool, CDS, Strasbourg, France.
The original description of the VizieR service was published in A\&AS 143, 23.


\end{document}